\documentclass[12pt,double]{article}

\usepackage{times}
\usepackage{graphicx} 
\usepackage{amsmath}
\usepackage{amsthm}
\usepackage{amssymb}
\usepackage{enumerate}
\usepackage{verbatim}
\usepackage{float}
\usepackage[margin=0.8in]{geometry}
\usepackage[list=true]{subcaption} 
\usepackage{multicol,ragged2e,setspace}
\usepackage{bm}
\usepackage[round]{natbib}

\usepackage{array}
\newcolumntype{H}{>{\setbox0=\hbox\bgroup}c<{\egroup}@{}}

\newcommand{\E}{\mathrm{E}}

\newcolumntype{H}{>{\setbox0=\hbox\bgroup}c<{\egroup}@{}}

\begin{document} 
\title{\textbf{Raking and Regression Calibration: Methods to Address Bias from Correlated Covariate and Time-to-Event Error}}
\author{}

\date{\vspace{-5ex}}

\maketitle

\begin{center}
Eric J. Oh$^{\footnotemark 1}$, Bryan E. Shepherd$^{2}$, Thomas Lumley$^{3}$, Pamela A. Shaw$^{1}$  \\
\vspace{0.1in}
$^1$University of Pennsylvania, Perelman School of Medicine \\
Department of Biostatistics, Epidemiology, and Informatics \\
\vspace{0.1in}
$^{2}$Vanderbilt University School of Medicine \\
Department of Biostatistics \\
\vspace{0.1in}
$^{3}$University of Auckland \\
Department of Statistics
\end{center}

\footnotetext[1]{Corresponding author: ericoh@pennmedicine.upenn.edu}

\doublespacing

\begin{abstract}
{Medical studies that depend on electronic health records (EHR) data are often subject to measurement error, as the data are not collected to support research questions under study. These data errors, if not accounted for in study analyses, can obscure or cause spurious associations between patient exposures and disease risk. Methodology to address covariate measurement error has been well developed; however, time-to-event error has also been shown to cause significant bias but methods to address it are relatively underdeveloped. More generally, it is possible to observe errors in both the covariate and the time-to-event outcome that are correlated. We propose regression calibration (RC) estimators to simultaneously address correlated error in the covariates and the censored event time. Although RC can perform well in many settings with covariate measurement error, it is biased for nonlinear regression models, such as the Cox model. Thus, we additionally propose raking estimators which are consistent estimators of the parameter defined by the population estimating equation. Raking can improve upon RC in certain settings with failure-time data, require no explicit modeling of the error structure, and can be utilized under outcome-dependent sampling designs. We discuss features of the underlying estimation problem that affect the degree of improvement the raking estimator has over the RC approach. Detailed simulation studies are presented to examine the performance of the proposed estimators under varying levels of signal, error, and censoring. The methodology is illustrated on observational EHR data on HIV outcomes from the Vanderbilt Comprehensive Care Clinic.} 
\end{abstract}

\section{Introduction}
\label{sec1}

 Biomedical research relies increasingly on electronic health records (EHR) data, either as the sole or supplemental source of data, due to the vast amount of data these resources contain and their relatively low cost compared to prospectively collected data. However, EHR data and other large cohort databases have been observed to be error-prone. These errors, if not accounted for in the data analysis, can bias associations of patient exposures and disease risk. There exists a large body of literature describing the impact of and methods to correct for covariate measurement error (\citealp{carroll2006measurement}); however, much less attention has been given to errors in the outcome. For linear models, independent random (classical) errors in the outcome variable do not bias regression estimates; however, errors correlated with either predictors in the model or errors in those predictors could bias associations. For non-linear models, even classical outcome errors can bias estimated associations of interest (\citealp{carroll2006measurement}). There are many examples in clinical research where the outcome of interest relies on an imprecisely measured event time. Researchers studying the epidemiology of chronic conditions may enroll subjects some time after an initial diagnosis, and so research questions focused on the timing of events post diagnosis may need to rely on patient recall or chart review of electronic medical records for the date of diagnosis, both of which are subject to error. Errors in the time origin can be systematic, as subject characteristics can influence the amount of error in recall. Methods to handle a misclassified outcome have been developed for binary outcomes (\citealp{magder1997logistic, edwards2013accounting, le2016evaluating}) and discrete failure time data (\citealp{meier2003discrete, magaret2008incorporating, hunsberger2010analysis}), where estimates of sensitivity and specificity can be incorporated into the bias correction. However, methods to handle errors in a continuous failure time have largely been ignored. 

Additionally, as more and more observational studies utilize data primarily collected for non-research purposes (e.g. administrative databases or electronic health records), it is increasingly common to have errors in both the outcome and exposures that are correlated. For example, in some observational studies of HIV/AIDS, the date of antiretroviral therapy (ART) initiation has been observed to have substantial errors (\citealp{shepherd2011accounting, duda2012measuring}). These errors can lead to errors in event times, defined as time since ART initiation, and errors in exposures of interest, such as CD4 count at ART initiation. Furthermore, certain types of records are often more likely to have errors (e.g. records from a particular study site), records with errors often tend to have errors across multiple variables, and the magnitude of these errors cannot be assumed uncorrelated. Ignoring correlated outcome and exposure errors could lead to positive or negative bias in estimates of regression parameters.   

In some settings, data errors can be corrected by retrospectively reviewing and validating medical records; however, this is expensive and time-consuming to do for a large number of records. Instead, we can perform data validation on a subset of selected records and use this information to correct estimates based on the larger, unvalidated dataset. In this manuscript, we propose regression calibration and raking estimators as two methods to correct the bias induced from such correlated errors by incorporating information learned in a validation subset to the large unvalidated dataset. 

Regression calibration (RC), introduced by \citet{prentice1982covariate}, is a method to address covariate measurement error that is widely used due to ease of implementation and good numerical performance in a broad range of settings. Although most RC methods assume measurement error in covariates only, \citet{2018arXiv181110147S} examined a way to apply RC to correlated errors in a covariate and a continuous outcome; to date these methods have not addressed correlated errors between failure time outcomes and exposures.

Raking is a method in survey sampling that makes use of auxiliary information available on the population to improve upon the Horvitz-Thompson (HT) estimator for regression parameters in two-phase designs. The HT estimator is known to be inefficient (\citealp{robins1994estimation}) but raking improves statistical efficiency, without changing the target of inference, by adjusting the standard HT weights by tuning them to auxiliary variables. Raking also takes advantage of the known sampling probabilities with validation studies such as those considered in this manuscript. These survey sampling ideas, while not new, have not been carefully studied in the measurement error setting. \citet{breslow2009improved} considered raking estimators for modeling case-cohort data with missing covariates. \citet{lumley2011connections} considered a raking estimator using simulated data in a covariate measurement error context with a validation subset. In this manuscript, we consider raking estimators for more general settings allowing for errors in the covariate and a time-to-event outcome, including misclassification, and discuss various possibilities for the auxiliary variables, how different choices affect the degree of improvement over the HT estimator, and ways to implement these methods using standard statistical software.

Our contributions in this manuscript are twofold. First, we develop regression calibration estimators to address both censored event time error alone and correlated covariate and censored event time errors together. To our knowledge, no RC estimators have been developed for these settings. Second, we develop raking estimators that are consistent and, in some settings, improve upon the RC estimators. These methods are important given the increased use of error-prone data in biomedical research and the paucity of methods that simultaneously handle errors in covariates and times-to-event. The rest of the paper proceeds as follows. We present our survival time model and the considered measurement error frameworks in Section 2. Sections 3 and 4 present the proposed regression calibration and raking methods, respectively. Section 5 compares the relative performance of the proposed estimators with simulation studies for various parameter settings and error distributions. In Section 6, we apply our methods to an HIV cohort and ascertain their robustness to misclassification. We conclude with a discussion in Section 7.

\section{Time-to-Event Model and Error Framework}\label{framework}

We consider the Cox proportional hazards model. Let $T_i$ and $C_i$,  be the failure time and right censoring time, respectively, for subjects $i=1,\ldots,n$ on a finite follow-up time interval, $[0,\tau]$. Define $U_i=\text{min}(T_i,C_i)$ and the corresponding failure indicator $\Delta_i = I(T_i \leq C_i)$. Let $Y_i(t) = I(U_i \geq t)$ and $N_i(t)=I(U_i \leq t, \Delta_i=1)$ denote the at-risk indicator and counting process for observed events, respectively. Let $X_i$ be a \emph{p}-dimensional vector of continuous covariates that are measured with error and $Z_i$ a \emph{q}-dimensional vector of precisely measured discrete and/or continuous covariates that may be correlated with $X_i$. We assume $C_i$ is independent of $T_i$ given $(X_i,Z_i)$ and that the data are i.i.d. Let the hazard rate for subject $i$ at time $t$ be given by $\lambda_i(t) = \lambda_0(t)\exp(\beta_X' X_i + \beta_Z' Z_i)$, where $\lambda_0(t)$ is an unspecified baseline hazard function. We consider $\beta_X$ to be the parameter(s) of interest, which is estimated by solving the partial likelihood score for $\beta = \left(\beta_X, \beta_Z\right)$.
\begin{equation}\label{cox_score}
\sum_{i=1}^n \int_0^{\tau} \left\{\left\{X_i,Z_i\right\}' - \frac{n^{-1}\sum_{j=1}^n Y_j(t)\left\{X_j,Z_j\right\}'\exp(\beta_X' X_j + \beta_Z' Z_j)}{n^{-1}\sum_{j=1}^n Y_j(t)\exp(\beta_X' X_j + \beta_Z' Z_j)}\right\} dN_i(t) = 0
\end{equation}

\subsection{Additive Measurement Error Structure}
\label{additive_error}

Oftentimes, errors seen in electronic health records data  or other datasets used for observational studies will not be simple random error and will depend on other variables in the dataset. For example, when the time-to-event error is due to a mismeasured time origin, this timing error can cause correlated errors in the baseline observations for exposures that are associated with the true survival outcome. In addition, errors induced in the exposures and censored time-to-event outcome can vary systematically with  subject characteristics that could make a subject's record more error-prone. Thus, we consider the error setting involving additive systematic and random error in both the covariates and time-to-event.

Instead of observing $(X,Z,U,\Delta)$, we observe $(X^{\star},Z, U^{\star},\Delta)$, where
\begin{align}
X^{\star} &= \alpha_0 + \alpha_1' X + \alpha_2' Z + \epsilon \\
U^{\star} &= U + \gamma_0 + \gamma_1' X + \gamma_2' Z + \nu = U + \omega.
\end{align}
Note that $X$ and $Z$ in the above formulation do not necessarily represent the full vector of covariates (e.g. some elements in the vectors $\alpha_1$, $\alpha_2$, $\gamma_1$, and $\gamma_2$ may be 0). We assume that $\epsilon$ and $\nu$ are mean 0 random variables with variance $\Sigma_{\epsilon\epsilon}$ and $\Sigma_{\nu\nu}$, respectively, and are independent of all other variables with the exception that we allow their covariance, $\Sigma_{\epsilon\nu}$, to be non-zero. We refer to this setting as the \textit{additive error structure}. In this setting the error in the observed censored failure time $U^*$ is a mistiming error but there are no errors in the event indicator $\Delta$.

\subsection{More General Error Structure}\label{misclassification_section}

We will see in the sections to follow that raking estimators, contrary to regression calibration estimators, do not require modeling the measurement error structure explicitly. Thus, we will also consider a more general error model that also involves a misspecified event. Whereas the additive error structure in Section \ref{additive_error} might be expected in scenarios involving only an error-prone baseline time (e.g. self-reported baseline time), the general error model relaxes this assumption to allow the timing of the failure, and thus the failure indicator, to be error-prone as well. Instead of observing $(X,Z,U,\Delta)$, one observes $(X^{\star},Z,U^{\star},\Delta^{\star})$, where errors in the event may be coming from both a mistiming error and also from misclassification of the event indicator. Note that with this error structure we also make no assumptions regarding the additivity of errors or their correlation with other variables. 

\subsection{Two-Phase Design}\label{two_phase_design}

We consider the two-phase design in which the true, error-free variables are measured retrospectively for a subsample of subjects at the second phase. Let $R_i$ be an indicator for whether subject $i=1,\ldots,n$ is selected to be in the second phase and let $0 < \pi_i \leq 1$ be their known sampling probability. In general, the sampling probabilities are known in validation studies based on observational data utilizing EHR, which are becoming increasingly common. This sampling scheme also accommodates scenarios where the subsample size is fixed (e.g. simple random sampling) and where the subsample size is random (e.g. Bernoulli sampling), as well as stratified designs (e.g. case-cohort). We assume that at phase one, the random variables $(X_i^{\star},Z_i, U_i^{\star},\Delta_i)$ [or $(X^{\star}_i,Z_i,U^{\star}_i,\Delta^{\star}_i)$ in a setting with misclassification] are observed for $n$ subjects as a random sample from the population. At phase two, $m < n$ subjects are selected from the phase one population according to the aforementioned sampling probability and the random variables $(X_i,U_i)$ [or $(X_i,U_i,\Delta_i)$] are additionally observed for those subjects. From this point on, we refer to the phase two subjects as the validation subset.

\section{Proposed Regression Calibration Methodology}\label{regcal}

In this section, we give a brief introduction to the original RC and risk set regression calibration (RSRC) methods for classical covariate measurement error and then develop their extensions for our considered error settings that include error in the censored outcome alone and correlated errors in the censored outcome and covariates. Under regularity conditions similar to those in \citet{andersen1982cox}, the RC and RSRC estimators developed in this section for error in the censored outcome and potentially correlated errors in the censored outcome and covariates are asymptotically normal, although not necessarily consistent for $\beta$. The proof is similar to that in the covariate error only setting, which was shown in \citet{wang1997regression}. For more detail see Appendix A of the Supplementary Materials.  

\subsection{Regression Calibration for Covariate Error}

\citet{prentice1982covariate} introduced the regression calibration method for the setting of Cox regression and classical measurement error in the covariate. \citet{shaw2012hazard} applied regression calibration for the covariate error structure assumed in Section \ref{additive_error}. The idea of regression calibration is to estimate the unobserved true variable with its expectation given the data. \citet{prentice1982covariate} showed that under the independent censoring assumption, the induced hazard function based on the error-prone data is given by $\lambda(t ; X^{\star},Z) = \lambda_0(t)\exp\left(\beta_Z' Z\right)\E\left(\exp\{\beta_X' X\} | X^{\star}, Z, U \geq t\right)$. He then showed that for rare events and moderate $\beta_X$, $\E\left(\exp\{\beta_X' X\} | X^{\star}, Z, U \geq t\right) \approx \exp\left(\beta_X' \E\left(X|X^{\star},Z\right)\right)$. $\E(X|X^{\star},Z)$ can be estimated using the following first order approximation
\begin{equation}\label{RC_x}
\E(X|X^{\star},Z ) = \mu_X + 
\begin{bmatrix}
\Sigma_{XX^{\star}} & \Sigma_{XZ}
\end{bmatrix}
\begin{bmatrix}
\Sigma_{X^{\star}X^{\star}} & \Sigma_{X^{\star}Z} \\
\Sigma_{ZX^{\star}} & \Sigma_{ZZ}
\end{bmatrix}^{-1}
\begin{bmatrix}
X^{\star}-\mu_{X^{\star}} \\
Z - \mu_{Z}
\end{bmatrix},
\end{equation}
where the validation subset is used to calculate the moments involving $X$ (see \citet{shaw2012hazard}). Define $\hat{X}=\E(X|X^{\star},Z ; \hat{\zeta}_x)$, where $\hat{\zeta}_x$ is the vector of nuisance parameters in (\ref{RC_x}) estimated from the data. $\hat{X}$ is then imputed for $X$ in the partial likelihood score (\ref{cox_score}) instead of the observed $X^{\star}$ to solve for $\beta$, which yields the corrected estimates (\citealp{shaw2012hazard}). Note, for simplicity we generally suppress the notation of the dependence of terms such as $\E(X|X^{\star},Z)$ on the nuisance parameter $\zeta_x$, unless it is important for clarity, such as to refer to its estimator $\E(X|X^{\star},Z ; \hat{\zeta}_x)$.

\subsection{Proposed Regression Calibration Extension for Time-to-Event Error}\label{RC_outcome}

Assume the time-to-event error structure in Section \ref{additive_error}, i.e., we observe $(X, Z, U^{\star},\Delta)$. Given the additivity of the outcome errors in (3), we can take the expectation of the censored event time, $U^{\star}$, given the observed covariates and rearrange to obtain $\E(U | X,Z) = \E(U^{\star}|X,Z) - \E\left(\omega|X,Z\right)$. We use $\E(\omega|X,Z)$ to correct $U^{\star}$ and then impute as our estimate of the true censored event time. Since the true $\E(\omega|X,Z)$ is unknown, we can estimate it using the following first order approximation 
\begin{equation}\label{RC_y}
\E(\omega|X,Z ; \zeta_{\omega}) = \mu_{\omega} + 
\begin{bmatrix}
\Sigma_{\omega X} & \Sigma_{\omega Z}
\end{bmatrix}
\begin{bmatrix}
\Sigma_{XX} & \Sigma_{XZ} \\
\Sigma_{ZX} & \Sigma_{ZZ}
\end{bmatrix}^{-1}
\begin{bmatrix}
X-\mu_{X} \\
Z - \mu_{Z}
\end{bmatrix},
\end{equation}
where the validation subset is used to calculate the moments involving $\omega$ and $\zeta_{\omega}$ is the vector of nuisance parameters in (\ref{RC_y}). Adjusting $U^{\star}$ to have the correct expectation gives us $\hat{U} = U^{\star} - \E(\omega|X,Z ; \hat{\zeta}_{\omega})$, which we use instead of $U^{\star}$ to solve the partial likelihood score (\ref{cox_score}) for the corrected $\beta$ estimates. 

\subsection{Proposed Regression Calibration Extension for Covariate and Time-to-Event Error}\label{RC_cov_outcome}

Assume the additive error structure for both $X^{\star}$ and $U^{\star}$ in Section \ref{additive_error}, i.e., we observe $(X^{\star}, Z, U^{\star}, \Delta)$. Given the additivity of the outcome errors in (2.3), we can take the expectation of the censored event time, $U^{\star}$, given the observed covariates and rearrange to obtain $\E(U | X^{\star},Z) = \E(U^{\star}|X^{\star},Z) - \E\left(\omega|X^{\star},Z\right)$. We use $\E(\omega|X^{\star},Z)$ to correct $U^{\star}$ and then impute as our estimate of the true censored event time. Due to the error-prone $X^{\star}$, we impute $\E(X|X^{\star},Z)$ for $X^{\star}$ as well, similar to \cite{prentice1982covariate}. Given that the true $\E(X|X^{\star},Z ; \zeta_x)$ is unknown, we estimate it using the same first order approximation described in Section 3.1. In addition, we propose to estimate $\E(\omega|X^{\star},Z ; \zeta_{\omega})$ using the same first order approximation described in Section 3.2 except using $X^{\star}$ instead of $X$, giving us $\hat{U} = U^{\star} - \E(\omega|X^{\star},Z ; \hat{\zeta}_{\omega})$ as the estimate of the true censored time-to-event. Thus, we impute $\hat{U}$ and $\hat{X}=\E(X|X^{\star},Z ; \hat{\zeta}_x)$ in the partial likelihood score (\ref{cox_score}) instead of the observed $U^{\star}$ and $X^{\star}$ and solve for $\beta$ to obtain our corrected estimates. 

\subsection{Proposed Risk Set Regression Calibration (RSRC) extension}\label{rsrc_section}

We also considered improving our regression calibration estimators by applying the idea of recalibrating the mismeasured covariate within each risk set developed by \citet{xie2001risk} for classical measurement error and extended to the covariate error model in Section 2.1 by \citet{shaw2012hazard}. Since the risk set membership likely depends on subject specific covariates whose distribution is changing over time, we may be able to obtain better RC estimates by performing the calibration at every risk set as events occur. In particular, this method was shown to decrease the bias significantly for the setting of covariate measurement error when the hazard ratio is quite large, a case in which ordinary RC has been observed to perform poorly. Specifically for covariate measurement error, the risk set regression calibration estimator solves the partial likelihood score (\ref{cox_score}) using $\hat{X}(t)$ instead of $X$, where $\hat{X}(t)$ is recalculated using RC at each event time using data from only those individuals still in the risk set at that event time. 

In the presence of time-to-event error, however, the necessary moments needed to estimate the conditional expectations in Sections \ref{RC_outcome} and \ref{RC_cov_outcome} at the $i^{th}$ individuals' censored event time will be incorrect due to the fact that the risk sets defined by $U^{\star}$ will not be the same as those defined by $U$, leading to biased estimates. Thus, to extend the RSRC idea to the settings of error in the censored outcome and correlated error in the covariate and censored outcome, we propose a two-stage RSRC estimator where the first stage involves obtaining the estimate $\hat{U}$ using ordinary RC. The second stage then assumes $\hat{U}$ is the observed event time instead of $U^{\star}$ and recalibrates $\hat{U}$ and $X^{\star}$ at risk sets defined by $\hat{U}$ using the methods described in Section \ref{RC_outcome} and Section \ref{RC_cov_outcome}.

\section{Proposed Generalized Raking Methodology}\label{raking}

In this section, we develop design-based estimators by applying generalized raking (raking for short) (\citealp{deville1992calibration,deville1993general}), which leverages the error-prone data available on the entire sample to improve the efficiency of consistent estimators calculated using the error-free validation subset. We give a brief overview of the general raking method and then propose our estimators for the correlated measurement error settings under consideration. Under suitable regularity conditions, the proposed raking estimators have been shown to be $\sqrt{n}$ consistent, asymptotically normal estimators of $\beta$ for all two-phase designs described in Section \ref{two_phase_design}. For the proof, see \citet{saegusa2013weighted}.

\subsection{Generalized Raking Overview}

Let $P_i(\beta)$ denote the population score equations for the true underlying Cox model with corresponding target parameter $\beta$, the log hazard ratio we would estimate if we had error-free data on the full cohort. Then the HT estimator of $\beta$ is given by the solution to $\sum_{i=1}^n \frac{R_i}{\pi_i} P_i(\beta) = 0$, which is known to be a consistent estimator of $\beta$. Consider $A_i$, a set of auxiliary variables that are available for everyone at phase one and are correlated with the phase two subsample variables. Raking estimators modify the design weights $w_{i, des}=\frac{1}{\pi_i}$ to new weights $w_{i, cal}=\frac{g_i}{\pi_i}$ such that they are as close as possible to $w_{i, des}$ while $\sum_{i=1}^n A_i$ is exactly estimated by the validation subset. Thus, given a distance measure $d(.,.)$, the objective is
\begin{align}
\text{minimize} &\sum_{i=1}^n R_i d\left(\frac{g_i}{\pi_i},\frac{1}{\pi_i}\right) \nonumber \\
\text{subject to} & \sum_{i=1}^n A_i = \sum_{i=1}^n R_i \frac{g_i}{\pi_i} A_i.
\end{align}
Note that the constraints above are known as the calibration equations. \citet{deville1993general} give several options for choosing the distance function, and the resulting constrained minimization problem can be solved to yield a solution for $g_i$. The generalized raking estimator is then defined as the solution to
\begin{equation}\label{raking_eqn}
\sum_{i=1}^n R_i \frac{g_i}{\pi_i} P_i(\beta) = 0.
\end{equation}

\subsection{Proposed Raking Estimators}\label{auxiliary_choice}

For our setting of the Cox model, we use the distance function $d(a,b) = a \log\left(\frac{a}{b}\right) + (b-a)$ in the objective function of (6) to ensure positive weights. Solving the constrained minimization problem for $g_i$ then yields $g_i=\exp\left(-\hat{\lambda}' A_i\right)$. After plugging in $g_i$ to the calibration equations, \citet{deville1992calibration} show that the solution for $\lambda$ satisfies 
\begin{equation*}\label{raking_lambda}
\hat{\lambda} = \hat{B}^{-1} \left(\sum_{i=1}^N \frac{R_i}{\pi_i} A_i - \sum_{i=1}^N A_i\right) + O_p(n^{-1}),    
\end{equation*}
where $\hat{B} = \sum_{i=1}^N \frac{R_i}{\pi_i} A_i' A_i$. Finally, we construct auxiliary variables, $A_i$, that yield efficient estimators. 

\citet{breslow2009improved} derived the asymptotic expansion for the solution to (\ref{raking_eqn}) and showed that the optimal auxiliary variable is given by $A_i^{\text{opt}} = \E(\tilde{\ell}_0(X_i,Z_i,U_i,\Delta_i)|V)$, where $\tilde{\ell}_0(X_i,Z_i,U_i,\Delta_i)$ denotes the efficient influence function contributions from the population model had the true outcome and covariates been observed for everyone in phase one and $V=(X^{\star},Z,U^{\star},\Delta)$ [or $(X^{\star},Z,U^{\star},\Delta^{\star})$ in a setting with misclassification]. However, calculating $A_i^{\text{opt}}$ involves a conditional distribution of unobserved variables and thus is generally not practically obtainable. \citet{kulich2004improving} proposed a ``plug in'' method that approximates this conditional expectation by using the influence functions from a model fit using phase one data. Specifically, they proposed to use the phase two data to fit models that impute the missing information from the phase one data only and then to obtain the influence functions from the desired model that uses imputed values in place of the missing data.  They further proposed using a \textit{dfbeta} type residual, which is readily available in statistical software, to estimate the influence function from the approximate model. We will propose two different imputations for the missing data, which will lead to two different choices of $A_i$ that approximate $A_i^{\text{opt}}$.

The first proposed approximation of $A_i^{\text{opt}}$ is given by $A_{N,i}=\tilde{\ell}_0(X^{\star}_i,Z_i,U^{\star}_i,\Delta_i)$, the influence function for the naive estimator that used the error prone data instead of the unobserved true values. One can estimate  $A_{N,i}$ empirically using
\begin{align*}
\tilde{\ell}_0(X^{\star}_i,Z_i,U^{\star}_i,\Delta_i) & \approx \Delta_i\left\{\left\{X^{\star}_i,Z_i\right\}' - \frac{S^{(1)\star}(\beta,t)}{S^{(0)\star}(\beta,t))}\right\} \\
&- \sum_{i=1}^n \int_0^{\tau} \frac{\exp(\beta_X' X^{\star}_i + \beta_Z' Z_i)}{S^{(0)\star}(\beta,t)} \left\{\left\{X^{\star}_i,Z_i\right\}' - \frac{S^{(1)\star}(\beta,t)}{S^{(0)\star}(\beta,t))}\right\} dN^{\star}_i(t),
\end{align*}
where $S^{(r)\star}(\beta,t) = n^{-1}\sum_{j=1}^n Y^{\star}_j(t)\left\{X^{\star}_j,Z_j\right\}^{'\otimes r}\exp(\beta_X' X^{\star}_j + \beta_Z' Z_j)$ ($a^{\otimes 1}$ is the vector $a$ and $a^{\otimes 0}$ is the scalar 1). For measurement error settings including an error-prone failure indicator, we approximate $A_i^{\text{opt}}$ with $A_{N,i}= \tilde{\ell}_0(X^{\star}_i,Z_i,U^{\star}_i,\Delta^{\star}_i)$.

The second proposed approximation of $A_i^{\text{opt}}$ is given by $A_{\text{RC},i}=\tilde{\ell}_0(\hat{X}_i(\hat{\zeta}_x),Z_i,\hat{U}_i(\hat{\zeta}_\omega),\Delta_i)$, i.e., the influence function for the target estimator that uses the calibrated estimates $(\hat{X}_i(\hat{\zeta}_x),\hat{U}_i(\hat{\zeta}))$ in place of the unobserved true data $(X_i,U_i)$. One can again use the empirical approximation  
\begin{align*}
\tilde{\ell}_0(\hat{X}_i(\hat{\zeta}_x),Z_i,\hat{U}_i(\hat{\zeta}_\omega),\Delta_i) & \approx  \Delta_i\left\{\left\{\hat{X}_i(\hat{\zeta}_x),Z_i\right\}' - \frac{\hat{S}^{(1)}(\beta,\hat{\zeta},t)}{\hat{S}^{(0)}(\beta,\hat{\zeta},t))}\right\} \\
&- \sum_{i=1}^n \int_0^{\tau} \frac{\exp(\beta_X' \hat{X}_i(\hat{\zeta}_x) + \beta_Z' Z_i)}{\hat{S}^{(0)}(\beta,\hat{\zeta},t)} \left\{\left\{\hat{X}_i(\hat{\zeta}_x),Z_i\right\}' - \frac{\hat{S}^{(1)}(\beta,\hat{\zeta},t)}{\hat{S}^{(0)}(\beta,\hat{\zeta},t))}\right\} d\hat{N}_i(t;\hat{\zeta}_{\omega}),
\end{align*}
where $\hat{S}^{(r)}(\beta,\hat{\zeta},t) = n^{-1}\sum_{j=1}^n \hat{Y}_j(t;\hat{\zeta}_{\omega})\left\{\hat{X}_j(\hat{\zeta}_x),Z_j\right\}^{'\otimes r}\exp(\beta_X' \hat{X}_j(\hat{\zeta}_x) + \beta_Z' Z_j)$ ($a^{\otimes 1}$ is the vector $a$ and $a^{\otimes 0}$ is the scalar 1).  For measurement error settings including an error-prone failure indicator, we approximate $A_i^{\text{opt}}$ with $A_{\text{RC},i} = \tilde{\ell}_0(\hat{X}_i(\hat{\zeta}_x),Z_i,\hat{U}_i(\hat{\zeta}_\omega),\Delta^{\star}_i)$. Thus, the two proposed raking estimators are:
\begin{enumerate}
\item Generalized raking naive (GRN): solution to (\ref{raking_eqn}) using $A_{N,i}$
\item Generalized raking regression calibration (GRRC): solution to (\ref{raking_eqn}) using $A_{\text{RC},i}$
\end{enumerate}
where both estimators utilize $g_i=\exp\left(-\hat{\lambda}' A_i\right)$.

The efficiency gain from the raking estimator over the HT estimator depends on the correlation between the auxiliary variables and the target variables. \citet{breslow2007weighted} showed that the variance of HT parameter estimates is the sum of the model-based variance due to sampling from an infinite population with no missing data and the design-based variance resulting from estimation of the unknown full cohort total of efficient influence function contributions. Thus, we consider $\tilde{\ell}_0(X_i,Z_i,U_i,\Delta_i)$ to be our target variables. We expect the regression calibration estimators to be less biased than the naive estimators and therefore conjecture that $A_{\text{RC}}$ would be more highly correlated with $A^{\text{opt}}$ than $A_N$. Note that in general, when the parameter of interest is a regression parameter, choosing the auxiliary variables to be the observed, error-prone variables will not improve efficiency. For more details, see Chapter 8 of \citet{lumley2011complex}. 

\subsection{Calculating Raking Estimators}\label{calc_raking}

Instead of explicitly calculating $A_{N,i}$ and $A_{\text{RC},i}$ with the influence function formulas given above, we propose to utilize standard software to calculate the $A_i$ so that practitioners may easily implement these methods.  In \textsf{R}, the influence functions can be approximated with negligible error as a \emph{dfbeta} type residual. Thus, the raking estimates can be computed as follows:
\begin{enumerate}
\item Fit a candidate Cox model using all phase one subjects.
\item Construct the auxiliary variables $A_i$ as imputed \emph{dfbetas} from the model fit in Step 1.
\item Estimate regression parameters $\beta$ using weights raked to $A_i$ by solving (\ref{raking_eqn}).
\end{enumerate}
For step one, we consider the naive Cox model using the error-prone data (GRN) and the regression calibration approach described in Section \ref{regcal} (GRRC). For step three, we utilize the \emph{survey} package by \citet{lumley2016package} in \textsf{R}, which provides standard software for obtaining raking estimates.

\section{Simulation Studies}

We examined the finite sample performance of our proposed RC, RSRC, GRRC, and GRN estimators through simulation for the error framework described in Section \ref{framework}. These four estimators were compared to those from the true model, a Cox proportional hazards regression model fit with the true covariates and event times, a naive Cox model fit with the error-prone covariates and/or error-prone censored event times, and the complete-case estimator using only the true covariates and event times in the validation subset. We note that all validation subsets were selected as simple random samples with known sampling probability, meaning the complete-case estimator is equivalent to the HT estimator. Following Section \ref{additive_error}, we considered the additive error structure with correlated covariate and time-to-event error. In addition to this case, we also considered the censored outcome error only setting. We further considered correlated covariate and censored outcome error under the special case where the covariates are only subject to random error, namely classical measurement error $\left((\alpha_0,\alpha_2)=\vec{0} ; \alpha_1=\vec{1}\right)$. In addition, we considered the general error structure described in Section \ref{misclassification_section}, where there exists errors in the time-to-event that result from mistiming as well as misclassification in addition to additive covariate error. We present $\%$ biases, average bootstrap standard errors (ASE) for the 4 proposed estimators or average model standard errors (ASE) for the naive and complete case estimators, empirical standard errors (ESE), mean square errors (MSE), and 95$\%$ coverage probabilities (CP) for varying values of the log hazard ratio $\beta_X$, $\%$ censoring, and error variances and covariances. We additionally present type 1 error results for $\beta_X=0$ and $\alpha = 0.05$. 

\subsection{Simulation Set-up}

All simulations were run 2000 times using $\textsf{R}$ version 3.4.2. The error-prone covariate X was generated as a standard normal distribution and the error-free covariate as $Z \sim \text{N}(2,1)$, with $\rho_{X,Z}=0.5$. We set the true log hazard ratios to be $\beta_X \in \{\log(1.5),\log(3)\}$, which we refer to as moderate and large, respectively, and $\beta_Z = \log(2)$. The true survival time $T$ was generated from an exponential distribution with rate equal to $\lambda_0 \exp(\beta_X X + \beta_Z Z)$, where $\lambda_0 = 0.1$. We then simulated $25\%$ and $75\%$ censoring, which we refer to as common and rare event settings, respectively, by generating separate random right censoring times for each $\beta_X$ to yield the desired $\%$ censored event times. Censoring times were generated as Uniform distributions with length $2$ and $0.4$ for each $\%$ censored time, respectively, to mimic studies of different lengths. For the error terms $\epsilon$ and $\nu$, we considered normal distributions with means 0, variances $\left(\Sigma_{\epsilon\epsilon} = \sigma^2_{\epsilon}, \Sigma_{\nu\nu} = \sigma^2_{\nu}\right) \in \{0.5, 1\}$, and $\left(\Sigma_{\epsilon\nu} = \sigma_{\epsilon\nu}\right) \in \left\{0.15, 0.3\right\}$, resulting in correlations ranging from $0.15$ to $0.60$. The error-prone covariate and censored event time were generated with parameters $(\alpha_0, \alpha_1, \alpha_2) = (0, 0.9, -0.2)$ and $(\gamma_0, \gamma_1, \gamma_2) = (\sigma_{\nu} \times 3, 0.2, -0.3)$. The choice of $\gamma_0$ is such that the error-prone time is a valid event time (i.e., greater than zero) with high probability. The few censored event times that were less than 0 were reflected across 0 to generate valid outcomes.

For the error terms $\epsilon$ and $\nu$, we also considered a mixture of a point mass at zero and a shifted gamma distribution with the same means and covariances as the normal distributions to determine the robustness of our methods to non-normality of errors. Note that while the RC and RSRC estimators are expected to be challenged by such departures from normality, the raking estimators are not affected by the structure of the measurement error other than by the strength of the correlation between the auxiliary variables and the target variables. The mixture probability was set to be $0.5$ for both covariate and outcome error. 

For the misclassification example, we set $\beta_X = \log(1.5)$, $\sigma^2_{\epsilon}=\sigma^2_{\nu}=0.5$, $\sigma_{\epsilon\nu}=0.15$, with normally distributed error terms and $75\%$ censoring. In addition, the sensitivity and specificity for $\Delta$ were set to $90\%$ by adding Bernoulli error $(p=0.10)$. For all simulations, we set the number of subjects to be $2000$ and selected the validation subsets as simple random samples of size 200, or $\pi_i=\pi=0.1$. The data example in Section 6 considers selecting the validation subsets using unequal sampling probabilities via outcome-dependent sampling. 

Standard errors for the RC, GRRC, and GRN estimates were obtained using the bootstrap method with bootstrap sampling stratified on the validation subset membership and using $300$ bootstrap samples. Note that while the raking estimators have known sandwich variance estimators for the asymptotic variance, we used the bootstrap to calculate standard errors and coverage probabilities (see Appendix B of the Supplementary Materials for an empirical comparison). The RSRC standard errors were also calculated similarly using the bootstrap; however, only $100$ bootstrap samples were utilized due to its computational burden. In addition, the RSRC estimators were recalibrated at deciles of the observed event times.  

\subsection{Simulation Results}

For all discussed tables, we observed that the naive estimates had very large bias with $95\%$ coverage hovering around $0\%$. In contrast, the complete case estimates were nearly unbiased for all settings discussed, but suffered from large standard errors, particularly for rare event settings when there were only a few subjects who had events in the validation subset. The coverage of the complete case estimates was near $95\%$ for all settings. In the discussion of simulation results to follow, we focus on the 4 proposed estimators and how their relative performance differed across settings.  

Table \ref{outcome_err_table} presents the relative performance for estimating $\beta_X$ in the presence of the time-to-event error described in Section \ref{additive_error} and no covariate error, with $\nu \sim \text{N}(0,\sigma^2_{\nu})$. The RC estimates had moderate to large bias ($-13\%$ to $-33\%$) and coverage ranging from $0.87$ to $0$, depending on if $\beta_X$ was moderate or large. We observed around a $50\%$ decrease in bias for the RSRC estimates compared to RC for moderate $\beta_X$ and common events and a range of $5-30\%$ bias reduction for other settings, with coverage around $87-93\%$ and $0\%$ for moderate and large $\beta_X$, respectively. The reduction in bias for the RSRC estimates resulted in a lower MSE for all settings except under moderate $\beta_X$ and rare events, a setting in which RC is known to perform well. Both raking estimates were nearly unbiased across all parameter settings, had uniformly lower standard errors than the complete case estimates, and had coverage near $95\%$. Interestingly, the performances of the GRRC and GRN estimators were virtually indistinguishable, with similar bias, standard errors, MSE, and coverage. Overall, RSRC had the lowest MSE for all moderate $\beta_X$ settings whereas the raking estimates had the lowest MSE for all large $\beta_X$ settings.  

Tables \ref{correlated_err_table1} and \ref{correlated_err_table2} consider the relative performance for estimating a moderate log hazard ratio in the setting of correlated additive errors in the outcome and covariate as described in Section \ref{additive_error} for normally distributed error terms and common and rare events, respectively. The RC estimates had relatively moderate bias ($-13\%$ to $-19\%$) and coverage ranging from $0.74$ to $0.92$. For common events, the RSRC estimates had around $50\%$ less bias than the RC estimates, whereas for rare events, they yielded only a small decrease in bias. Even in these more complex error settings, both raking estimates remained nearly unbiased, had lower standard errors than the complete case estimates, and maintained coverage around $95\%$ across varying error variances and covariances. We noticed that for all parameter settings, the GRRC and GRN estimators were again nearly indistinguishable. Overall for the common event settings, the RSRC estimates had the lowest MSE when the error variances were both $0.5$; otherwise, the raking estimates had the lowest MSE for all other settings. For the rare event settings, the RC estimates had the lowest MSE across all variance and covariance settings.

We present the relative performance for estimating a larger log hazard ratio, keeping other parameters the same as in Tables \ref{correlated_err_table1} and \ref{correlated_err_table2}, in Table \ref{correlated_err_table3} and Supplementary Materials Table 1 in Appendix C. Both the RC and RSRC estimates had large bias, ranging from $-31\%$ to $-37\%$ and $-23\%$ to $-32\%$, respectively, as well as coverage $50\%$ or below. Again, both raking estimates remained nearly unbiased, had lower standard errors than the complete case estimates, and maintained coverage around $95\%$ across varying error variances and covariances, with the GRRC and GRN estimates indistinguishable. Across all error settings, the raking estimates had the lowest MSE. 

Table \ref{type1_table} presents the type 1 error, ASE, ESE, and MSE when $\beta_X=0$ in the presence of correlated, additive measurement error in the outcome and covariate $X$ with normally distributed errors. For both levels of censoring, the type 1 error of the RC and RSRC estimates ranged from $0.044$ to $0.059$ and the raking estimates were around $0.042$ and $0.046$ for common and rare events, respectively. It is of note that the type 1 error for the naive estimator is $1$ for both levels of censoring, meaning the null hypothesis was falsely rejected in every simulation run. 

Results for $\beta_Z$, for the settings presented in Tables \ref{outcome_err_table}-\ref{correlated_err_table3}, are presented in Tables 2-5 of Appendix C in the Supplementary Materials. The conclusions for this parameter were similar to those of $\beta_X$; however, the raking estimates had the lowest MSE across more settings. Tables 6-8 in Appendix D of the Supplementary Materials present simulation results for $\beta_X$ in a setting where the covariates are only subject to classical measurement error, keeping all other settings the same as Tables \ref{correlated_err_table1}-\ref{correlated_err_table3}. Results are similar to those presented above.

We consider the relative performance for when the error distributions were generated as a mixture of a point mass at 0 and shifted gamma distribution, with settings otherwise the same as those in Tables \ref{outcome_err_table}-\ref{correlated_err_table3}, in Tables 9-12 of Appendix E in the Supplementary Materials. The RC and RSRC estimators were challenged by such departures from normality, with generally more bias and higher MSE, while the raking estimators remained unbiased with lower MSE.

Table 13 in Appendix F of the Supplementary Materials considers the relative performance of the estimators in the presence of misclassification errors in addition to the correlated additive errors in the time-to-event and covariate $X$, as described in Section \ref{misclassification_section}. The RC and RSRC estimates had very large bias and coverage between $61\%$ and $68\%$ as these methods were not developed to directly handle misclassification. As expected, the GRRC and GRN estimates were nearly unbiased because the raking estimators do not depend on the structure of the measurement error. Overall, the raking estimators had the lowest MSE in this more complex error setting.  

\section{Data Example}

We applied the four proposed methods to electronic health records data from a large HIV clinic, the Vanderbilt Comprehensive Care Clinic (VCCC). The VCCC is an outpatient clinic that provides care to HIV patients and collects clinical data over time that is electronically recorded by nurses and physicians (\citealp{lemly2009race}). The VCCC fully validated all key variables for all records, resulting in an unvalidated, error-prone dataset and a fully validated dataset that we consider to be correct. Thus, this observational cohort is ideal for directly assessing the relative performance of the proposed regression calibration and raking estimators compared to the naive and HT estimators. Note that the naive estimator was calculated using only the unvalidated dataset as if the validated dataset did not exist. In addition, the HT estimator was calculated using a subsample of the fully validated dataset. Throughout this example, we considered the estimates from the fully validated dataset to be the ``truth'' and defined these as the parameters of interest. In addition, all considerations of bias were relative to these target parameters. We considered two different failure time outcomes of interest: time from the start of antiretroviral therapy (ART) to the time of virologic failure and to the time of first AIDS defining event (ADE). For the former analysis, virologic failure was defined as an HIV-RNA count greater than or equal to 400 copies/mL and patients were censored at the last available test date after ART initiation. The HIV-RNA assay, and hence time at virologic failure was largely free of errors, whereas the time at ART start was error-prone, corresponding to errors in $U$. The ADE outcome was defined as the first opportunistic infection (OI) and patients were censored at age of death if it occurred or last available test date after ART initiation. For this failure time, both time of ART initiation and time at first ADE were error-prone, corresponding to errors in $U$ and $\Delta$. We studied the association between the outcomes of interest and the CD4 count and age at ART initiation. Since date of ART initiation was error prone, CD4 and age at ART initiation may also have errors. Appendix G of the Supplementary Materials provides detail on the eligibility criteria and statistics for the covariate and time-to-event error for both analyses.

The analysis of the virologic failure outcome included 1863 patients with moderate censoring rates of $46.1\%$ and $47.2\%$ in the unvalidated and validated dataset, respectively. We observed highly (slightly) skewed error in CD4 count at ART start (observed event times) and very small amounts of misclassification. The validation subset was selected as a simple random sample of $20\%$, resulting in 373 patients. For this sampling design, the HT estimator is equivalent to the complete case estimator. The hazard ratios and their corresponding confidence intervals comparing the estimators are displayed graphically in the first row of Figure \ref{vccc_plot} and shown in Table 14 in Appendix H of the Supplementary Materials. We note that the standard errors for all estimators (including the true, naive, and HT) were calculated using the bootstrap with 300 replicates, which were somewhat larger than the model SEs likely due to a lack of fit of the Cox model.  The RSRC estimators were recalibrated at vigintiles of the observed event times. For this analysis, there was little bias in the naive estimators of a 100 cell/$\text{mm}^3$ increase in CD4 count at ART initiation and 10 year increase of age at ART initiation ($1.87\%$ and $2.17\%$, respectively). For both covariates, RC and RSRC provided very minimal improvements in bias, albeit with slightly wider confidence intervals. Small bias notwithstanding, we noticed that both the GRRC and GRN estimators had smaller bias compared to the naive estimator and had narrower confidence intervals than the HT estimator. The GRRC and GRN estimators had very little differentiating them, similar to what was observed in the simulations.

The analysis of the ADE outcome included 1595 patients with very high censoring rates of $84.5\%$ and $93.8\%$ in the unvalidated and validated dataset, respectively. We observed highly (slightly) skewed error in CD4 count at ART start (observed event times) and a misclassification rate of $11\%$ that was largely due to false positives $(\text{positive predictive value}=35\%)$. While the RC and RSRC methods developed in this paper do not explicitly handle misclassification, we were nevertheless interested in seeing how they would perform in this real data scenario in comparison to the raking methods that can handle misclassification. Due to ADE being a rare event, we utilized a case-cohort sampling scheme to select the validation subset. Specifically, we selected a simple random sample of $7\%$, or 112 patients, from the full error-prone data and then added the remaining 227 subjects classified as cases by the error-prone ADE indicator to the validation subset. Note that due to the biased sampling scheme of the case-cohort design, the estimates of the conditional expectations involved in the RC and RSRC estimators cannot be calculated in the same manner as under simple random sampling. Thus, we used IPW least squares to estimate the conditional expectations for RC, RSRC, and GRRC (step one of calculating raking estimates as detailed in Section \ref{calc_raking}) . The hazard ratios and their corresponding confidence intervals comparing the estimators are displayed graphically in the second row of Figure \ref{vccc_plot} and shown in Table 14 in Appendix H of the Supplementary Materials. The standard errors for all estimators were again calculated using the bootstrap with 300 replicates. We noticed significantly more bias in the naive estimators of a 100 cell/$\text{mm}^3$ increase in CD4 count at ART initiation and 10 year increase of age at CD4 count measurement ($31.44\%$ and $31.2\%$, respectively). In fact, the naive point estimate for age was in the wrong direction compared to the true estimate, yielding anticonservative bias. The RC and RSRC estimators provided little to no bias improvement for both covariates. However, the GRRC and GRN estimates were both nearly unbiased with narrower confidence intervals than those of the HT estimator. Again, we noticed that the GRRC and GRN estimators gave similar estimates, with GRRC (GRN) having narrower confidence intervals for the CD4 (age) hazard ratios. In this analysis, we noticed huge improvements in bias from the GRRC and GRN estimators compared to the naive estimators and decreased standard errors compared to the HT estimator even in the presence of appreciable misclassification, which the RC and RSRC estimators could not handle.

The \textsf{R} package \textsf{RRCME} at https://github.com/ericoh17/RRCME implements our methods on a simulated data set that mimics the structure of the VCCC data. Additionally, Appendix I of the Supplementary Materials contains code that implements the RC and GRN estimators for this simulated data to demonstrate ease of application of these estimators.

\section{Discussion}

Data collected primarily for non-research purposes, such as those from administrative databases or EHR, can have errors in both the outcome and exposures of interest, which can be correlated. Using EHR data from the VCCC HIV cohort, we observed that Cox regression models using the unvalidated dataset as compared to the fully validated dataset resulted in a 3-fold underestimation of the CD4 hazard ratio for ADE and overestimation of the age hazard ratio in the wrong direction such that the null hypothesis of a unit hazard ratio was nearly rejected. Spurious associations driven by such unvalidated outcomes and exposures can misdirect clinical researchers and can be harmful to patients down the line. Even when variables are reviewed and validated for a subset of the records, the additional information gained from these validation procedures are not often utilized in estimation. 

The existing literature does not adequately address such complex error across multiple variables; in particular, the timing error in the censored failure time outcome. In this article, we developed four different estimators that incorporate an internal validation subset in the analysis to try to obtain unbiased and efficient estimates. The RC and RSRC estimators approximate the true model by estimating the true outcome and/or exposure given the unvalidated data and information on the error structure from the validation subset. This approximation lacks consistency in most cases for nonlinear models and the RC and RSRC estimators can have appreciable bias for some error settings. However, in settings with a modest hazard ratio and rare events, RC outperformed the other estimators with respect to having the lowest MSE. RSRC had the lowest MSE for settings with a modest hazard ratio and common events under only censored outcome error and for settings with a modest hazard ratio, common events, and small error variance under correlated outcome and covariate error. The proposed regression calibration methods were considered for the proportional hazards model; however, we expect they would work quite well more generally in accelerated failure time models where an additive error structure is assumed. In fact, some forms of error in the outcome will bias the proportional hazards parameter but not the acceleration parameter (\citealp{oh2018considerations}). 

The generalized raking estimators are consistent whenever the design-weighted complete case estimating equations (e.g. HT estimator) yields consistent estimators; they use influence functions based on the unvalidated data as auxiliary variables to improve efficiency over the complete case estimator and can be used under outcome-dependent sampling. The raking estimators are not sensitive to the measurement error structure, which is in contrast to the RC and RSRC estimators that can perform poorly when the error structure is not correctly specified. In particular, we noticed in our data example and simulations that in the presence of misclassification as well as timing errors, GRRC and GRN yield nearly unbiased estimates while RC and RSRC are substantially biased. Generally, the raking estimators performed well, with little small sample bias and, in most cases, the smallest MSE. The raking estimators had large efficiency gains in settings with a large hazard ratio as well those with a modest hazard ratio, common events, and large error variances. For all settings considered, GRRC and GRN performed similarly. GRN has the added advantage that it can be applied with standard statistical software, e.g. the survey package in \textsf{R} (\citealp{lumley2016package}).

As noted above, the performance of the GRRC and GRN estimators was virtually identical, contrary to our hypothesis that the GRRC estimates would be more efficient than those of GRN. This result was unknown for previous applications of raking (\citealp{breslow2009improved}, \citealp{lumley2011connections}) and in fact goes against their recommendation to build imputation models for the partially missing variables. For the setting of only classical covariate measurement error and no time-to-event error,  we derived (not shown) that the influence functions for Cox regression using $X{^\star}$ versus $\hat{X}$ are scalar multiples of each other. Thus, the solutions to (\ref{raking_eqn}) under both auxiliary variables are equivalent. For the more complex error settings considered in this paper (Sections \ref{additive_error}, \ref{misclassification_section}), an explicit characterization of the relationship between the two auxiliary variables is more difficult, but we hypothesize that an approximation of a similar type holds for the settings studied. 

The motivating example for this paper was to develop methods where there were only errors in the failure time outcome but not in the failure indicator. We additionally considered methods, namely GRRC and GRN, that are able to address more general error structures. We believe future research investigating RC methods to directly correct for misclassification resulting from time-to-event error would be worthwhile. In addition, while theory demonstrates that generalized raking estimators are consistent, we noticed that the small sample bias (and efficiency) can depend on the specific validation subsample. Developing optimal subsampling schemes to maximize efficiency would not only improve the complete case analysis, but also increase the efficiency gains of the raking estimators and is an area of future work.

\section*{Acknowledgements}
We would like to thank Timothy Sterling, MD and the co-investigators of the Vanderbilt Comprehensive Care Clinic (VCCC) for use of their data. This work was supported by a Patient Centered Outcomes Research Institute (PCORI) Award [R-1609-36207] and the U.S. National Institutes of Health (NIH) [R01-AI131771, P30-AI110527, R01-AI093234, U01-AI069923, and U01-AI069918]. The statements in this manuscript are solely the responsibility of the authors and do not necessarily represent the views of PCORI or NIH.

\bibliographystyle{plainnat}
\bibliography{raking_RC}

\newpage

\begin{figure}[H]
\begin{center}
\includegraphics[scale=0.475]{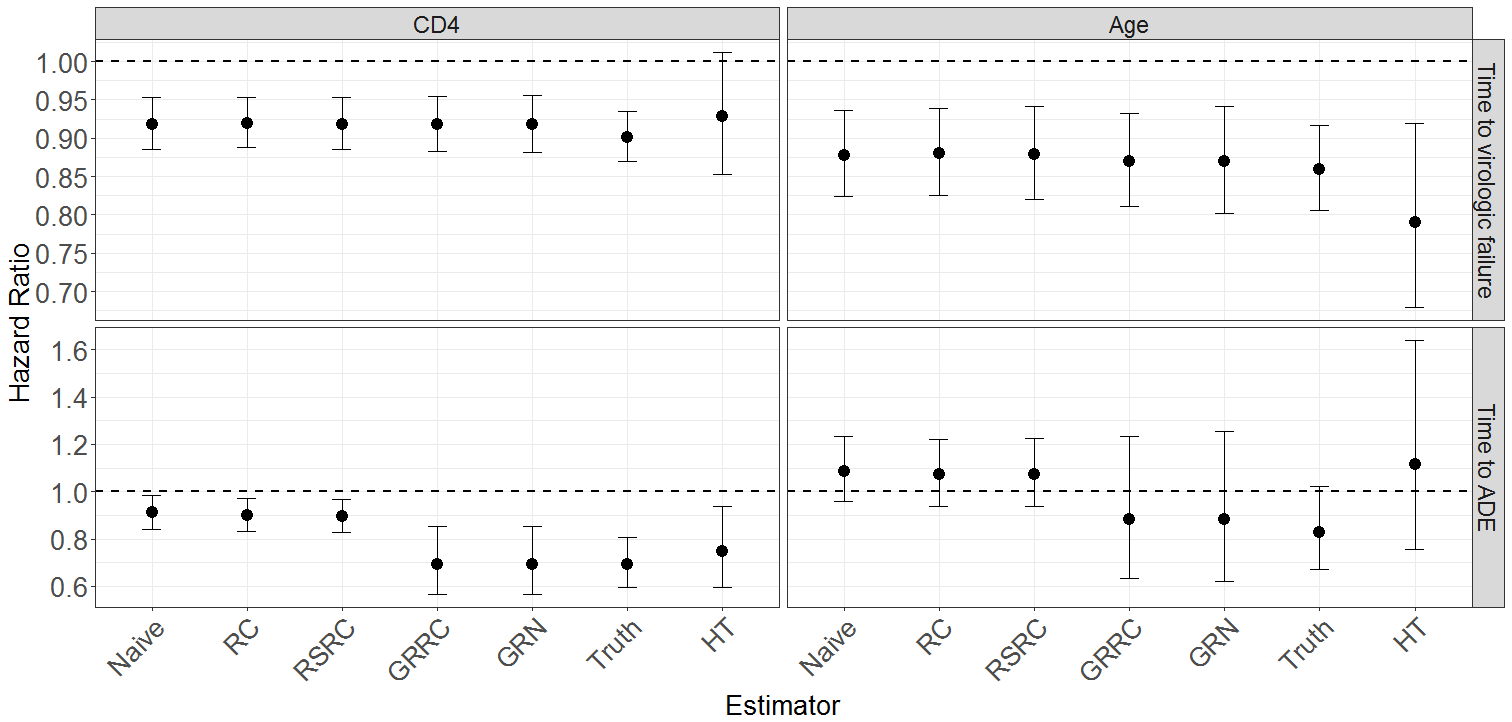}
\caption{The hazard ratios and their corresponding $95\%$ confidence intervals (CI) for a 100 cell/$\text{mm}^3$ increase in CD4 count at ART initiation and 10 year increase in age at CD4 count measurement. Estimates and their CIs are calculated using the bootstrap for the Regression Calibration (RC), Risk Set Regression Calibration (RSRC), Generalized Raking Regression Calibration (GRRC), and Generalized Raking Naive (GRN) estimators.}
\label{vccc_plot}
\end{center}
\end{figure}

\begin{table}[H]
\renewcommand{\thetable}{\arabic{table}}
\caption{Simulation results for $\beta_X$ under additive measurement error only in the outcome with normally distributed error and 25 and 75$\%$ censoring for the true event time. For $2000$ simulated data sets, the bias, average bootstrap standard error (ASE) for the 4 proposed estimators, average model standard error (ASE) for naive and complete case, empirical standard error (ESE), mean squared error (MSE), and 95$\%$ coverage probabilities (CP) are presented.}
\label{outcome_err_table}
\centering
\scalebox{0.71}{
  \begin{tabular}{cccccccccH}
  $\%$ Censoring & $\beta_X$ & $\sigma^{2}_{\nu}$ & Method & $\%$ Bias & ASE & ESE & MSE & CP & Power\\
  \hline
  25 & log(1.5) & & True &-0.025 & 0.030 & 0.031 & 0.001 & 0.947 & 1\\
  \\
  & & 0.5 & RC & -12.677 & 0.042 & 0.043 & 0.004& 0.752 & 1\\
  & & & RSRC & -5.056 & 0.048 & 0.050 & 0.003& 0.928 & 1\\
  & & & GRRC & 0.074 & 0.059 & 0.058 & 0.003& 0.957 & 1\\
  & & & GRN &0.271 & 0.060 & 0.059 & 0.003& 0.958 & 1\\
  & & & Naive &-37.562 & 0.030 & 0.031 & 0.024& 0.002 & 1\\
  & & & Complete &0.321 & 0.098 & 0.098 & 0.010 & 0.952 & 0.989\\
  & & 1 & RC & -18.522 & 0.046 & 0.047 & 0.008& 0.624 & 1\\
  & & & RSRC  & -7.991 & 0.055 & 0.056 & 0.004& 0.910 & 1\\
  & & & GRRC & -0.025 & 0.066 & 0.065 & 0.004& 0.956 & 1\\
  & & & GRN &0.074 & 0.066 & 0.065 & 0.004& 0.958 & 1\\
  & & & Naive &-40.891 & 0.030 & 0.030 & 0.028& 0.000 & 1\\
  & & & Complete &0.321 & 0.098 & 0.098 & 0.010& 0.954 & 0.988\\
  \\
  & log(3) & & True & 0.046 & 0.037 & 0.036 & 0.001 & 0.951 \\
  \\
  & & 0.5 & RC & -26.879 & 0.054 & 0.056 & 0.090 & 0.001 & 1\\
  & & & RSRC & -19.188 & 0.060 & 0.063 & 0.048& 0.070 & 1\\
  & & & GRRC & -0.983 & 0.103 & 0.102 & 0.010& 0.938 & 1\\
  & & & GRN &-1.010 & 0.104 & 0.104 & 0.011& 0.939 & 1\\
  & & & Naive &-37.347 & 0.031 & 0.040 & 0.170 & 0.000 & 1\\
  & & & Complete &0.819 & 0.118 & 0.118 & 0.014& 0.954 & 1\\
  & & 1 & RC & -33.042 & 0.056 & 0.058 & 0.135& 0.000 & 1\\
  & & & RSRC & -23.466 & 0.065 & 0.067 & 0.071& 0.027 & 1\\
  & & & GRRC & -0.883 & 0.108 & 0.105 & 0.011& 0.940 & 0.999\\
  & & & GRN &-0.847 & 0.108 & 0.106 & 0.011& 0.942 & 1\\
  & & & Naive &-41.88 & 0.030 & 0.039 & 0.213& 0.000 & 1\\
  & & & Complete &0.819 & 0.118 & 0.118 & 0.014& 0.955 & 1\\
  \\
  75 & log(1.5) & & True & 0.074 & 0.054 & 0.054 & 0.003& 0.948 & 1\\
  \\
  & & 0.5 & RC & -15.340 & 0.079 & 0.080 & 0.010& 0.872 & 0.988\\
  & & & RSRC & -12.874 & 0.087 & 0.089 & 0.011& 0.898 & 0.978\\
  & & & GRRC & -0.099 & 0.113 & 0.112 & 0.012& 0.957 & 0.953\\
  & & & GRN &0.543 & 0.116 & 0.117 & 0.014& 0.955 & 0.946\\
  & & & Naive &-69.204 & 0.054 & 0.055 & 0.082& 0.000 & 0.631\\
  & & & Complete &0.444 & 0.176 & 0.182 & 0.033& 0.950 & 0.63\\
  & & 1 & RC & -17.338 & 0.081 & 0.084 & 0.012& 0.845 & 0.982\\
  & & & RSRC & -15.488 & 0.089 & 0.092 & 0.012& 0.873 & 0.967\\
  & & & GRRC & -0.444 & 0.118 & 0.118 & 0.014& 0.952 & 0.938\\
  & & & GRN &0.247 & 0.120 & 0.121 & 0.015& 0.953 & 0.93\\
  & & & Naive &-57.638 & 0.054 & 0.056 & 0.058& 0.016 & 0.881\\
  & & & Complete &-0.099 & 0.177 & 0.182 & 0.033& 0.946 & 0.624\\
  \\
  & log(3) & & True & 0.118 & 0.058 & 0.059 & 0.003 & 0.950 \\
  \\
  & & 0.5 & RC & -31.030 & 0.085 & 0.088 & 0.124& 0.024 & 1\\
  & & & RSRC & -28.827 & 0.094 & 0.097 & 0.110& 0.087 & 1\\
  & & & GRRC & -0.901 & 0.166 & 0.163 & 0.027& 0.951 & 1\\
  & & & GRN &-0.446 & 0.168 & 0.175 & 0.031& 0.950 & 1\\
  & & & Naive &-52.357 & 0.053 & 0.062 & 0.335& 0.000 & 1\\
  & & & Complete &1.912 & 0.191 & 0.197 & 0.039& 0.946 & 1\\
  & & 1 & RC & -33.060 & 0.087 & 0.091 & 0.140& 0.024 & 1\\
  & & & RSRC & -31.567 & 0.095 & 0.099 & 0.130& 0.055 & 1\\
  & & & GRRC & -0.774 & 0.171 & 0.170 & 0.029& 0.940 & 1\\
  & & & GRN &-0.501 & 0.171 & 0.172 & 0.030& 0.942 & 1\\
  & & & Naive &-48.680 & 0.053 & 0.061 & 0.290& 0.000 & 1\\
  & & & Complete &1.930 & 0.193 & 0.202 & 0.041& 0.946 & 1\\
  \end{tabular}
}
\end{table}

\begin{table}[H]
\renewcommand{\thetable}{\arabic{table}}
\caption{Simulation results for $\beta_X=\log 1.5$ under correlated, additive measurement error in the outcome and covariate $X$ with normally distributed error and 25$\%$ censoring for the true event time. For $2000$ simulated data sets, the bias, average bootstrap standard error (ASE) for the 4 proposed estimators, average model standard error (ASE) for naive and complete case, empirical standard error (ESE), mean squared error (MSE), and 95$\%$ coverage probabilities (CP) are presented.}
\label{correlated_err_table1}
\centering
\scalebox{0.71}{
  \begin{tabular}{ccccccccccH}
  
  $\beta_X$ & $\sigma^{2}_{\nu}$ & $\sigma^2_{\epsilon}$ & $\sigma_{\nu,\epsilon}$ & Method & $\%$ Bias & ASE & ESE & MSE & CP & Power \\
  \hline
  log(1.5) & & & & True &-0.025 & 0.030 & 0.031 & 0.001 & 0.947 & 1\\
  \\
  & 0.5 & 0.5 & 0.15 & RC & -13.762 & 0.059 & 0.059 & 0.007& 0.804 & 1\\
  & & & & RSRC & -6.338 & 0.070 & 0.068 & 0.005& 0.922 & 1\\
  & & & & GRRC & 0.173 & 0.083 & 0.084 & 0.007& 0.947 & 0.998\\
  & & & & GRN &0.345 & 0.083 & 0.084 & 0.007& 0.946 & 0.998\\
  & & & & Naive &-79.760 & 0.024 & 0.025 & 0.105& 0.000 & 0.92\\
  & & & & Complete &0.321 & 0.098 & 0.098 & 0.010& 0.952 & 0.989\\
  \\
  & & & 0.30 & RC & -13.491 & 0.060 & 0.060 & 0.007& 0.813 & 1\\
  & & & & RSRC & -6.116 & 0.071 & 0.069 & 0.005& 0.928 & 1\\
  & & & & GRRC & 0.296 & 0.083 & 0.084 & 0.007& 0.947 & 0.998\\
  & & & & GRN & 0.567 & 0.083 & 0.084 & 0.007& 0.945 & 0.998\\
  & & & & Naive & -97.024 & 0.024 & 0.025 & 0.155& 0.000 & 0.094\\
  & & & & Complete &0.173 & 0.098 & 0.099 & 0.010& 0.954 & 0.987\\
  \\
  & & 1 & 0.15 & RC & -13.836 & 0.072 & 0.071 & 0.008& 0.843 & 1\\
  & & & & RSRC & -7.054 & 0.084 & 0.083 & 0.008& 0.922 & 0.996\\
  & & & & GRRC & 0.049 & 0.089 & 0.090 & 0.008& 0.948 & 0.996\\
  & & & & GRN & 0.148 & 0.089 & 0.090 & 0.008& 0.952 & 0.996\\
  & & & & Naive & -86.099 & 0.020 & 0.020 & 0.122& 0.000 & 0.804\\
  & & & & Complete &0.271 & 0.098 & 0.098 & 0.010& 0.952 & 0.989\\
  \\
  & & & 0.30 & RC & -13.639 & 0.073 & 0.072 & 0.008& 0.845 & 1\\
  & & & & RSRC & -6.955 & 0.086 & 0.084 & 0.008& 0.914 & 0.997\\
  & & & & GRRC & 0.074 & 0.089 & 0.090 & 0.008& 0.947 & 0.996\\
  & & & & GRN & 0.271 & 0.089 & 0.089 & 0.008& 0.945 & 0.996\\
  & & & & Naive & -97.912 & 0.020 & 0.020 & 0.158& 0.000 & 0.082\\
  & & & & Complete &0.222 & 0.098 & 0.098 & 0.010& 0.957 & 0.99\\
  \\
  & 1 & 0.5 & 0.15 & RC & -19.237 & 0.065 & 0.065 & 0.010& 0.746 & 1\\
  & & & & RSRC & -9.520 & 0.078 & 0.076 & 0.007& 0.902 & 0.997\\
  & & & & GRRC & 0.123 & 0.085 & 0.086 & 0.007& 0.944 & 0.996\\
  & & & & GRN &0.247 & 0.085 & 0.086 & 0.007& 0.944 & 0.996\\
  & & & & Naive &-79.686 & 0.024 & 0.025 & 0.105& 0.000 & 0.922\\
  & & & & Complete &0.321 & 0.098 & 0.098 & 0.010& 0.954 & 0.988\\
  \\
  & & & 0.30 & RC & -19.311 & 0.066 & 0.066 & 0.010& 0.743 & 1\\
  & & & & RSRC & -9.693 & 0.079 & 0.077 & 0.008& 0.903 & 0.998\\
  & & & & GRRC & 0.148 & 0.085 & 0.086 & 0.007& 0.945 & 0.996\\
  & & & & GRN & 0.345 & 0.085 & 0.085 & 0.007& 0.946 & 0.998\\
  & & & & Naive & -95.027 & 0.024 & 0.025 & 0.149& 0.000 & 0.146\\
  & & & & Complete &0.173 & 0.098 & 0.098 & 0.010& 0.955 & 0.986\\
  \\
  & & 1 & 0.15 & RC & -19.213 & 0.079 & 0.079 & 0.012& 0.801 & 0.993\\
  & & & & RSRC & -10.235 & 0.095 & 0.092 & 0.010& 0.908 & 0.975\\
  & & & & GRRC & -0.025 & 0.090 & 0.092 & 0.008& 0.945 & 0.994\\
  & & & & GRN & 0.074 & 0.090 & 0.091 & 0.008& 0.946 & 0.994\\
  & & & & Naive & -86.049 & 0.020 & 0.020 & 0.122& 0.000 & 0.812\\
  & & & & Complete &0.148 & 0.098 & 0.099 & 0.010& 0.952 & 0.986\\
  \\
  & & & 0.30 & RC & -19.213 & 0.080 & 0.080 & 0.012& 0.798 & 0.994\\
  & & & & RSRC & -10.580 & 0.096 & 0.093 & 0.010& 0.902 & 0.979\\
  & & & & GRRC & 0.123 & 0.090 & 0.091 & 0.008& 0.947 & 0.996\\
  & & & & GRN & 0.247 & 0.090 & 0.091 & 0.008& 0.948 & 0.996\\
  & & & & Naive & -96.556 & 0.020 & 0.020 & 0.154& 0.000 & 0.112\\
  & & & & Complete &0.321 & 0.098 & 0.098 & 0.010& 0.953 & 0.989\\
  \end{tabular}
}
\end{table}

\begin{table}[H]
\renewcommand{\thetable}{\arabic{table}}
\caption{Simulation results for $\beta_X=\log 1.5$ under correlated, additive measurement error in the outcome and covariate $X$ with normally distributed error and 75$\%$ censoring for the true event time. For $2000$ simulated data sets, the bias, average bootstrap standard error (ASE) for the 4 proposed estimators, average model standard error (ASE) for naive and complete case, empirical standard error (ESE), mean squared error (MSE), and 95$\%$ coverage probabilities (CP) are presented.}
\label{correlated_err_table2}
\centering
\scalebox{0.71}{
  \begin{tabular}{ccccccccccH}
  
  $\beta_X$ & $\sigma^{2}_{\nu}$ & $\sigma^2_{\epsilon}$ & $\sigma_{\nu,\epsilon}$ & Method & $\%$ Bias & ASE & ESE & MSE & CP & Power \\
  \hline
  log(1.5) & & & & True & 0.074 & 0.054 & 0.054 & 0.003 & 0.948 & 1\\
  \\
  & 0.5 & 0.5 & 0.15 & RC & -15.143 & 0.109 & 0.108 & 0.015& 0.906 & 0.895\\
  & & & & RSRC & -12.677 & 0.120 & 0.120 & 0.017& 0.925 & 0.854\\
  & & & & GRRC & 0.222 & 0.154 & 0.153 & 0.023& 0.955 & 0.772\\
  & & & & GRN &0.987 & 0.156 & 0.156 & 0.024& 0.956 & 0.763\\
  & & & & Naive &-120.208 & 0.046 & 0.046 & 0.240& 0.000 & 0.43\\
  & & & & Complete &0.444 & 0.176 & 0.182 & 0.033& 0.950 & 0.63\\
  \\
  & & & 0.30 & RC & -14.477 & 0.109 & 0.108 & 0.015& 0.900 & 0.914\\
  & & & & RSRC  & -11.715 & 0.121 & 0.119 & 0.016& 0.922 & 0.869\\
  & & & & GRRC & 0.099 & 0.154 & 0.152 & 0.023& 0.954 & 0.774\\
  & & & & GRN & 1.406 & 0.154 & 0.154 & 0.024& 0.954 & 0.776\\
  & & & & Naive & -167.043 & 0.048 & 0.049 & 0.461& 0.000 & 1\\
  & & & & Complete &0.444 & 0.177 & 0.183 & 0.034& 0.948 & 0.626\\
  \\
  & & 1 & 0.15 & RC & -14.896 & 0.134 & 0.131 & 0.021& 0.920 & 0.754\\
  & & & & RSRC & -13.047 & 0.146 & 0.146 & 0.024& 0.931 & 0.69\\
  & & & & GRRC & -0.099 & 0.166 & 0.164 & 0.027& 0.962 & 0.686\\
  & & & & GRN & 0.271 & 0.168 & 0.166 & 0.028& 0.958 & 0.683\\
  & & & & Naive & -113.623 & 0.038 & 0.038 & 0.214& 0.000 & 0.316\\
  & & & & Complete &0.271 & 0.177 & 0.183 & 0.034& 0.952 & 0.628\\
  \\
  & & & 0.30 & RC & -14.650 & 0.133 & 0.131 & 0.021& 0.922 & 0.76\\
  & & & & RSRC & -12.381 & 0.146 & 0.145 & 0.024& 0.936 & 0.702\\
  & & & & GRRC & 0.839 & 0.166 & 0.164 & 0.027& 0.958 & 0.704\\
  & & & & GRN & 1.430 & 0.168 & 0.167 & 0.028& 0.956 & 0.703\\
  & & & & Naive & -143.465 & 0.039 & 0.039 & 0.340& 0.000 & 0.994\\
  & & & & Complete &1.208 & 0.177 & 0.182 & 0.033& 0.948 & 0.627\\
  \\
  & 1 & 0.5 & 0.15 & RC & -16.993 & 0.113 & 0.114 & 0.018& 0.890 & 0.856\\
  & & & & RSRC & -15.316 & 0.123 & 0.123 & 0.019& 0.907 & 0.808\\
  & & & & GRRC & -0.370 & 0.156 & 0.155 & 0.024& 0.954 & 0.748\\
  & & & & GRN &0.444 & 0.158 & 0.157 & 0.024& 0.952 & 0.748\\
  & & & & Naive &-102.228 & 0.045 & 0.046 & 0.174& 0.000 & 0.058\\
  & & & & Complete &-0.099 & 0.177 & 0.182 & 0.033& 0.946 & 0.624\\
  \\
  & & & 0.30 & RC & -17.264 & 0.113 & 0.112 & 0.017& 0.892 & 0.862\\
  & & & & RSRC & -15.464 & 0.124 & 0.124 & 0.019& 0.904 & 0.805\\
  & & & & GRRC & -0.222 & 0.155 & 0.154 & 0.024& 0.956 & 0.75\\
  & & & & GRN & 0.814 & 0.156 & 0.155 & 0.024& 0.958 & 0.764\\
  & & & & Naive & -132.613 & 0.046 & 0.046 & 0.291& 0.000 & 0.822\\
  & & & & Complete &0.296 & 0.176 & 0.182 & 0.033& 0.950 & 0.636\\
  \\
  & & 1 & 0.15 & RC & -17.091 & 0.138 & 0.136 & 0.023& 0.918 & 0.702\\
  & & & & RSRC & -15.562 & 0.150 & 0.152 & 0.027& 0.916 & 0.631\\
  & & & & GRRC & -0.222 & 0.166 & 0.165 & 0.027& 0.957 & 0.7\\
  & & & & GRN & 0.123 & 0.168 & 0.167 & 0.028& 0.955 & 0.684\\
  & & & & Naive & -101.587 & 0.037 & 0.038 & 0.171& 0.000 & 0.06\\
  & & & & Complete &-0.074 & 0.176 & 0.182 & 0.033& 0.948 & 0.628\\
  \\
  & & & 0.30 & RC & -17.042 & 0.138 & 0.135 & 0.023& 0.916 & 0.696\\
  & & & & RSRC & -15.291 & 0.151 & 0.151 & 0.027& 0.916 & 0.639\\
  & & & & GRRC & 0.123 & 0.167 & 0.165 & 0.027& 0.954 & 0.691\\
  & & & & GRN & 0.814 & 0.169 & 0.167 & 0.028& 0.952 & 0.69\\
  & & & & Naive & -121.86 & 0.038 & 0.038 & 0.246& 0.000 & 0.65\\
  & & & & Complete &0.617 & 0.177 & 0.180 & 0.032& 0.954 & 0.638\\
  \end{tabular}
}
\end{table}

\begin{table}[H]
\renewcommand{\thetable}{\arabic{table}}
\caption{Simulation results for $\beta_X=\log 3$ under correlated, additive measurement error in the outcome and covariate $X$ with normally distributed error and 25$\%$ censoring for the true event time. For $2000$ simulated data sets, the bias, average bootstrap standard error (ASE) for the 4 proposed estimators, average model standard error (ASE) for naive and complete case, empirical standard error (ESE), mean squared error (MSE), and 95$\%$ coverage probabilities (CP) are presented.}
\label{correlated_err_table3}
\centering
\scalebox{0.695}{
  \begin{tabular}{ccccccccccH}
  
  $\beta_X$ & $\sigma^{2}_{\nu}$ & $\sigma^2_{\epsilon}$ & $\sigma_{\nu,\epsilon}$ & Method & $\%$ Bias & ASE & ESE & MSE & CP & Power \\
  \hline
  log(3) & & & & True & 0.055 & 0.037 & 0.036 & 0.001 & 0.952 & 1\\
  \\
  & 0.5 & 0.5 & 0.15 & RC & -31.239 & 0.077 & 0.077 & 0.124& 0.026 & 1\\
  & & & & RSRC & -23.038 & 0.092 & 0.092 & 0.072& 0.239 & 1\\
  & & & & GRRC & 0.337 & 0.113 & 0.112 & 0.012& 0.950 & 1\\
  & & & & GRN &0.346 & 0.112 & 0.111 & 0.012& 0.950 & 1\\
  & & & & Naive &-70.243 & 0.025 & 0.027 & 0.596& 0.000 & 1\\
  & & & & Complete &0.819 & 0.118 & 0.118 & 0.014& 0.954 & 1\\
  \\
  & & & 0.30 & RC & -31.904 & 0.079 & 0.080 & 0.129& 0.030 & 1\\
  & & & & RSRC & -23.102 & 0.097 & 0.096 & 0.074& 0.274 & 1\\
  & & & & GRRC & 0.410 & 0.113 & 0.111 & 0.012& 0.952 & 1\\
  & & & & GRN & 0.473 & 0.112 & 0.111 & 0.012& 0.954 & 1\\
  & & & & Naive & -76.842 & 0.024 & 0.026 & 0.713& 0.000 & 1\\
  & & & & Complete &0.810 & 0.118 & 0.118 & 0.014& 0.955 & 1\\
  \\
  & & 1 & 0.15 & RC & -31.895 & 0.094 & 0.093 & 0.132& 0.086 & 1\\
  & & & & RSRC & -24.394 & 0.111 & 0.110 & 0.084& 0.329 & 1\\
  & & & & GRRC & 0.373 & 0.116 & 0.115 & 0.013& 0.954 & 1\\
  & & & & GRN & 0.410 & 0.116 & 0.114 & 0.013& 0.952 & 1\\
  & & & & Naive & -79.473 & 0.020 & 0.022 & 0.763& 0.000 & 1\\
  & & & & Complete &0.719 & 0.118 & 0.118 & 0.014& 0.956 & 1\\
  \\
  & & & 0.30 & RC & -32.359 & 0.096 & 0.095 & 0.135& 0.092 & 1\\
  & & & & RSRC & -24.540 & 0.115 & 0.113 & 0.086& 0.351 & 1\\
  & & & & GRRC & 0.391 & 0.116 & 0.114 & 0.013& 0.957 & 1\\
  & & & & GRN & 0.455 & 0.115 & 0.114 & 0.013& 0.954 & 1\\
  & & & & Naive & -83.888 & 0.020 & 0.021 & 0.850& 0.000 & 1\\
  & & & & Complete &0.737 & 0.118 & 0.118 & 0.014& 0.956 & 1\\
  \\
  & 1 & 0.5 & 0.15 & RC & -35.900 & 0.079 & 0.079 & 0.162& 0.014 & 1\\
  & & & & RSRC & -26.916 & 0.095 & 0.094 & 0.096& 0.163 & 1\\
  & & & & GRRC & 0.328 & 0.114 & 0.112 & 0.013& 0.950 & 1\\
  & & & & GRN &0.337 & 0.114 & 0.112 & 0.013& 0.951 & 1\\
  & & & & Naive &-71.372 & 0.025 & 0.027 & 0.616& 0.000 & 1\\
  & & & & Complete &0.819 & 0.118 & 0.118 & 0.014& 0.955 & 1\\
  \\
  & & & 0.30 & RC & -36.528 & 0.080 & 0.081 & 0.168& 0.014 & 1\\
  & & & & RSRC & -27.334 & 0.098 & 0.097 & 0.100& 0.181 & 1\\
  & & & & GRRC & 0.337 & 0.114 & 0.112 & 0.013& 0.949 & 1\\
  & & & & GRN & 0.364 & 0.114 & 0.112 & 0.012& 0.954 & 1\\
  & & & & Naive & -76.997 & 0.024 & 0.026 & 0.716& 0.000 & 1\\
  & & & & Complete &0.728 & 0.118 & 0.118 & 0.014& 0.956 & 1\\
  \\
  & & 1 & 0.15 & RC & -36.246 & 0.096 & 0.096 & 0.168& 0.052 & 1\\
  & & & & RSRC & -28.409 & 0.114 & 0.113 & 0.110& 0.253 & 1\\
  & & & & GRRC & 0.391 & 0.117 & 0.115 & 0.013& 0.950 & 1\\
  & & & & GRN & 0.401 & 0.116 & 0.115 & 0.013& 0.950 & 1\\
  & & & & Naive & -80.256 & 0.020 & 0.022 & 0.778& 0.000 & 1\\
  & & & & Complete &0.755 & 0.118 & 0.118 & 0.014& 0.952 & 1\\
  \\
  & & & 0.30 & RC & -36.674 & 0.098 & 0.097 & 0.172& 0.056 & 1\\
  & & & & RSRC & -28.754 & 0.116 & 0.115 & 0.113& 0.264 & 1\\
  & & & & GRRC & 0.428 & 0.117 & 0.114 & 0.013& 0.952 & 1\\
  & & & & GRN & 0.446 & 0.116 & 0.114 & 0.013& 0.954 & 1\\
  & & & & Naive & -84.015 & 0.020 & 0.021 & 0.852& 0.000 & 1\\
  & & & & Complete &0.746 & 0.118 & 0.118 & 0.014& 0.954 & 1\\
  \\
  \end{tabular}
}
\end{table}

\begin{table}[H]
\renewcommand{\thetable}{\arabic{table}}
\caption{Type 1 error results for $\beta_X=0$ under correlated, additive measurement error in the outcome and covariates with normally distributed error and 25 and 75$\%$ censoring for the true event time. For $2000$ simulated data sets, the type 1 error, average bootstrap standard error (ASE) for the 4 proposed estimators, average model standard error (ASE) for naive and complete case, empirical standard error (ESE), and mean squared error (MSE) are presented. }
\label{type1_table}
\centering
\scalebox{0.75}{
\begin{tabular}{ccccccccc}
$\%$ Censoring & $\sigma^{2}_{\nu}$ & $\sigma^2_{\epsilon}$ & $\sigma_{\nu,\epsilon}$ & Method & Type 1 Error & ASE & ESE & MSE \\
\hline
25 & 0.5 & 0.5 & 0.15 & RC & 0.044 & 0.054 & 0.054 & 0.003 \\
& & & & RSRC & 0.050 & 0.063 & 0.062 & 0.004 \\
& & & & GRRC & 0.043 & 0.077 & 0.075 & 0.006 \\
& & & & GRN & 0.042 & 0.078 & 0.075 & 0.006 \\
& & & & Naive & 1.000 & 0.025 & 0.026 & 0.019 \\
& & & & Complete & 0.049 & 0.097 & 0.097 & 0.010 \\
\\
75 & 0.5 & 0.5 & 0.15 & RC & 0.050 & 0.102 & 0.102 & 0.010 \\
& & & & RSRC & 0.059 & 0.112 & 0.116 & 0.014 \\
& & & & GRRC & 0.046 & 0.141 & 0.141 & 0.020 \\
& & & & GRN & 0.046 & 0.143 & 0.143 & 0.021 \\
& & & & Naive & 1.000 & 0.045 & 0.047 & 0.080 \\
& & & & Complete & 0.056 & 0.170 & 0.178 & 0.032 \\
\end{tabular}
}
\end{table}

\pagebreak
\begin{center}
\textbf{\Large Supplementary Materials for ``Raking and Regression Calibration: Methods to Address Bias from Correlated Covariate and Time-to-Event Error''}
\end{center}

\begin{center}
Eric J. Oh$^{\star 1}$, Bryan E. Shepherd$^{2}$, Thomas Lumley$^{3}$, Pamela A. Shaw$^{1}$  \\
\vspace{0.1in}
$^1$University of Pennsylvania, Perelman School of Medicine \\
Department of Biostatistics, Epidemiology, and Informatics \\
\vspace{0.1in}
$^{2}$Vanderbilt University School of Medicine \\
Department of Biostatistics \\
\vspace{0.1in}
$^{3}$University of Auckland \\
Department of Statistics
\end{center}

\setcounter{equation}{0}
\setcounter{figure}{0}
\setcounter{table}{0}
\makeatletter
\renewcommand{\theequation}{S\arabic{equation}}
\renewcommand{\thefigure}{S\arabic{figure}}
\renewcommand{\bibnumfmt}[1]{[S#1]}
\renewcommand{\citenumfont}[1]{S#1}

\section*{Appendix A: Asymptotic theory for RC and RSRC estimators}

First, we consider the RC extension for covariate and time-to-event error in Section 3.3. The RC estimator in this setting, $\hat{\beta}_{RC}$, is found by solving the score equation
\begin{equation*}
S_{RC}(\beta, \hat{\zeta}) = \sum_{i=1}^n \int_0^{\tau} \left\{\left\{\hat{X}_i(\hat{\zeta}_x),Z_i\right\}'-\frac{S^{(1)}(\beta,\hat{\zeta}, t)}{S^{(0)}(\beta,\hat{\zeta},t)}\right\} d\hat{N}_i(t;\hat{\zeta}_{\omega}) = 0
\end{equation*}
where $S^{(r)}(\beta,\hat{\zeta},t) = n^{-1}\sum_{j=1}^n \hat{Y}_j(t;\hat{\zeta}_{\omega})\left\{\hat{X}_j(\hat{\zeta}_x),Z_j\right\}^{'\otimes r}\exp(\beta_X' \hat{X}_j(\hat{\zeta}_x) + \beta_Z' Z_j)$ ($a^{\otimes 1}$ is the vector $a$ and $a^{\otimes 0}$ is the scalar 1), and $\left\{\hat{U}(\hat{\zeta}_{\omega}), \hat{X}(\hat{\zeta}_x)\right\}$ are as given in Section 3.3. Throughout this section, we assume that $(1)$ $(N_i, Y_i, X_i, Z_i)$ are i.i.d; $(2)$ there exists a finite constant $\tau>0$ such that $P(U \geq \tau) > 0$; $(3)$ $\int_0^{\tau} \lambda_0(u) du < \infty$; and $(4)$ $\frac{m}{n} \rightarrow p \in (0,1)$. Define $\beta^{\star}$ as the solution to $\E\left\{S_{RC}(\beta,\zeta_0)\right\}=0$, which is generally not the same as $\beta$. First, we consider consistency for $\beta^{\star}$ and asymptotic normality for the solution to $S_{RC}(\beta,\zeta_0)$, where $\zeta_0=(\zeta_{x0},\zeta_{\omega 0})$ is the true nuisance parameter vector. Then $S_{RC}(\beta,\zeta_0)$, which is based on the standard Cox partial score equation, and thus concave, will have a unique, consistent solution, namely $\beta^{\star}$, under mild regularity conditions (see \citealp{andersen1982cox}). To establish asymptotic normality, we additionally define $\theta^{\star} = (\beta^{\star},\zeta_0)$ and assume that $(5)$ $\frac{\partial}{\partial \theta}S_{RC}(\theta)$ exists and is continuous and bounded for $\theta \in \mathcal{N}(\theta^{\star})$, a compact neighborhood of $\theta^{\star}$; $(6)$ $\frac{\partial}{\partial \theta}S_{RC}(\theta)$ converges to its limit $\E\left\{\frac{\partial}{\partial \theta}S_{RC}(\theta)\right\}$ uniformly in $\mathcal{N}(\theta^{\star})$; $(7)$ $\E\left\{\frac{\partial}{\partial \theta}S_{RC}(\theta)\right\}$ is nonsingular at $\theta^{\star}$; and $(8)$ $\E\left[\text{sup}_{\theta \in \mathcal{N}(\theta^{\star})} \left\{\left\{\hat{X}_j(\hat{\zeta}_x),Z_j\right\}\exp(\beta_X' \hat{X}_j(\hat{\zeta}_x) + \beta_Z' Z_j)\right\}^2\right] < \infty$. The techniques of \citet{andersen1982cox} can then be used to establish asymptotic normality of the solution to $S_{RC}(\beta,\zeta_0)$. Next, the solution to $S_{RC}(\beta, \hat{\zeta})$, where $\hat{\zeta}$ is our plug-in moment estimator for $\zeta$, can be shown to be consistent and asymptotically normal using Theorem 5.31 in \citet{van1998asymptotic}. The theorem additionally requires that $S_{RC}(\beta,\hat{\zeta})$ be Donsker in $\mathcal{N}(\theta^{\star})$. It is well known that the usual Cox score equation is Donsker and given that $\hat{\zeta}$ is a finite dimensional moment estimator, the estimating equations we solve to estimate the nuisance parameters are Donsker as well. $\hat{X}$ and $\hat{U}$ are Lipschitz transformations of $X$ and $U$ involving estimators from a Donsker class of functions, so it follows from Example 19.20 in \citet{van1998asymptotic} that $S_{RC}(\beta,\hat{\zeta})$ is Donsker.

The arguments above apply to show consistency and asymptotic normality of $\hat{\beta}_{RC}$ from Section 3.2 for time-to-event error only by utilizing the true $X$ instead of $\hat{X}$. Similary, the asymptotic properties of the RSRC estimators from Section 3.4 follow as well due to the fact that we recalibrate a fixed, finite number of times. This results in a finite number of Lipschitz transformations and thus a Donsker class of estimating equations. 

\section*{Appendix B: Empirical comparison of sandwich and bootstrap variances for raking estimators}

We used the bootstrap to calculate standard errors for the raking estimators due to the fact that we noticed coverage probabilities in some settings under $95\%$ using the sandwich variance estimators. For example, in an independent simulation with settings $\beta_X=\log(3)$, $\sigma^2_{\nu}=0.5$, $\sigma^2_{\epsilon}=1$, $\sigma_{\nu,\epsilon}=0.15$, and $25\%$ censoring, the coverage of GRRC was $0.9376$ using the sandwich estimator and $0.9524$ using the bootstrap. Note that Monte Carlo error cannot explain this undercoverage as the number of simulation runs was $2500$, resulting in a $95\%$ confidence interval of $0.95 \pm 1.96\sqrt{\frac{(0.95)(0.05)}{2500}}$, or $(0.9415,0.9585)$, which does not include $0.9376$. The coverage of GRN under the same settings was extremely similar.

\newpage

\section*{Appendix C: Additive error tables}

\begin{table}[H]
\renewcommand{\thetable}{\arabic{table}}
\caption{Simulation results for $\beta_X$ under correlated, additive measurement error in the outcome and covariate $X$ with $\beta_X=\log 3$, normally distributed error, and 75$\%$ censoring for the true event time. For $2000$ simulated data sets, the bias, average bootstrap standard error (ASE) for the 4 proposed estimators, average model standard error (ASE) for naive and complete case, empirical standard error (ESE), mean squared error (MSE), and 95$\%$ coverage probabilities (CP) are presented.}
\centering
\scalebox{0.68}{
\begin{tabular}{ccccccccccH}
  
  $\beta$ & $\sigma^{2}_{\nu}$ & $\sigma^2_{\epsilon}$ & $\sigma_{\nu,\epsilon}$ & Method & $\%$ Bias & ASE & ESE & MSE & CP & Power \\
  \hline
  log(3) & & & & True & 0.146 & 0.058 & 0.058 & 0.003 & 0.949 & 1\\
  \\
  & 0.5 & 0.5 & 0.15 & RC & -31.540 & 0.121 & 0.121 & 0.135& 0.198 & 1\\
  & & & & RSRC & -28.836 & 0.135 & 0.134 & 0.118& 0.339 & 0.999\\
  & & & & GRRC & 0.819 & 0.187 & 0.182 & 0.033& 0.960 & 1\\
  & & & & GRN &1.129 & 0.188 & 0.183 & 0.034& 0.958 & 1\\
  & & & & Naive &-86.163 & 0.044 & 0.047 & 0.898& 0.000 & 0.926\\
  & & & & Complete &1.912 & 0.191 & 0.197 & 0.039& 0.946 & 1\\
  \\
  & & & 0.30 & RC & -31.567 & 0.122 & 0.121 & 0.135& 0.214 & 1\\
  & & & & RSRC & -28.627 & 0.136 & 0.133 & 0.117& 0.356 & 1\\
  & & & & GRRC & 0.792 & 0.188 & 0.186 & 0.035& 0.955 & 1\\
  & & & & GRN & 1.329 & 0.187 & 0.187 & 0.035& 0.953 & 1\\
  & & & & Naive & -102.639 & 0.046 & 0.047 & 1.274& 0.000 & 0.093\\
  & & & & Complete &1.766 & 0.192 & 0.202 & 0.041& 0.940 & 1\\
  \\
  & & 1 & 0.15 & RC & -31.294 & 0.149 & 0.148 & 0.140 & 0.357 & 1\\
  & & & & RSRC & -28.827 & 0.166 & 0.164 & 0.127& 0.506 & 0.999\\
  & & & & GRRC & 1.283 & 0.194 & 0.191 & 0.037& 0.952 & 1\\
  & & & & GRN & 1.420 & 0.196 & 0.192 & 0.037& 0.954 & 1\\
  & & & & Naive & -90.669 & 0.036 & 0.038 & 0.994& 0.000 & 0.798\\
  & & & & Complete &1.957 & 0.192 & 0.200 & 0.041& 0.941 & 1\\
  \\
  & & & 0.30 & RC & -31.431 & 0.150 & 0.148 & 0.141& 0.354 & 1\\
  & & & & RSRC & -28.754 & 0.167 & 0.166 & 0.127& 0.492 & 0.998\\
  & & & & GRRC & 1.238 & 0.194 & 0.193 & 0.037& 0.958 & 1\\
  & & & & GRN & 1.611 & 0.196 & 0.192 & 0.037& 0.958 & 1\\
  & & & & Naive & -101.719 & 0.038 & 0.039 & 1.250& 0.000 & 0.086\\
  & & & & Complete &1.839 & 0.192 & 0.202 & 0.041& 0.942 & 1\\
  \\
  & 1 & 0.5 & 0.15 & RC & -33.415 & 0.123 & 0.124 & 0.150& 0.178 & 1\\
  & & & & RSRC & -31.695 & 0.137 & 0.135 & 0.139& 0.288 & 0.999\\
  & & & & GRRC & 0.847 & 0.190 & 0.187 & 0.035& 0.954 & 1\\
  & & & & GRN &1.174 & 0.190 & 0.188 & 0.036& 0.950 & 1\\
  & & & & Naive &-79.646 & 0.044 & 0.046 & 0.768& 0.000 & 0.998\\
  & & & & Complete &1.930 & 0.193 & 0.202 & 0.041& 0.946 & 1\\
  \\
  & & & 0.30 & RC & -33.652 & 0.124 & 0.123 & 0.152& 0.178 & 1\\
  & & & & RSRC & -31.494 & 0.138 & 0.138 & 0.139& 0.303 & 0.999\\
  & & & & GRRC & 0.874 & 0.188 & 0.186 & 0.034& 0.958 & 1\\
  & & & & GRN & 1.302 & 0.188 & 0.186 & 0.035& 0.956 & 0.999\\
  & & & & Naive & -90.541 & 0.045 & 0.046 & 0.992& 0.000 & 0.634\\
  & & & & Complete &1.866 & 0.192 & 0.201 & 0.041& 0.948 & 1\\
  \\
  & & 1 & 0.15 & RC & -33.378 & 0.152 & 0.151 & 0.157& 0.328 & 1\\
  & & & & RSRC & -31.804 & 11.643 & 0.166 & 0.149& 0.438 & 0.996\\
  & & & & GRRC & 1.129 & 0.195 & 0.193 & 0.037& 0.954 & 1\\
  & & & & GRN & 1.311 & 0.196 & 0.193 & 0.038& 0.952 & 0.999\\
  & & & & Naive & -86.191 & 0.036 & 0.038 & 0.898& 0.000 & 0.988\\
  & & & & Complete &1.866 & 0.192 & 0.201 & 0.041& 0.946 & 1\\
  \\
  & & & 0.30 & RC & -33.533 & 0.153 & 0.151 & 0.159& 0.328 & 1\\
  & & & & RSRC & -32.04 & 3.224 & 0.164 & 0.151& 0.439 & 0.992\\
  & & & & GRRC & 1.202 & 0.194 & 0.191 & 0.037& 0.951 & 1\\
  & & & & GRN & 1.538 & 0.195 & 0.192 & 0.037& 0.952 & 1\\
  & & & & Naive & -93.700 & 0.036 & 0.038 & 1.061& 0.000 & 0.464\\
  & & & & Complete &1.893 & 0.192 & 0.200 & 0.040& 0.944 & 1\\
  \\
  \end{tabular}
}
\end{table}

\begin{table}[H]
\renewcommand{\thetable}{\arabic{table}}
\caption{Simulation results for $\beta_Z$ under additive measurement error only in the outcome with normally distributed error and 25 and 75$\%$ censoring for the true event time. For $2000$ simulated data sets, the bias, average bootstrap standard error (ASE) for the 4 proposed estimators, average model standard error (ASE) for naive and complete case, empirical standard error (ESE), mean squared error (MSE), and 95$\%$ coverage probabilities (CP) are presented.}
\label{outcome_err_table_Z}
\centering
\scalebox{0.71}{
  \begin{tabular}{cccccccccH}
  
  $\%$ Censoring & $\beta_X$ & $\sigma^{2}_{\nu}$ & Method & $\%$ Bias & ASE & ESE & MSE & CP & Power\\
  \hline
  25 & log(1.5) & & True & 0.072 & 0.032 & 0.033 & 0.001 & 0.949 & 1\\
  \\
  & & 0.5 & RC & -12.523 & 0.044 & 0.043 & 0.009& 0.493 & 1\\
  & & & RSRC & -4.891 & 0.051 & 0.052 & 0.004& 0.884 & 1\\
  & & & GRRC & 0.115 & 0.066 & 0.065 & 0.004& 0.956 & 1\\
  & & & GRN &-0.014 & 0.066 & 0.065 & 0.004& 0.958 & 1\\
  & & & Naive &12.003 & 0.033 & 0.034 & 0.008& 0.294 & 1\\
  & & & Complete &1.428 & 0.104 & 0.105 & 0.011& 0.956 & 1\\
  & & 1 & RC & -18.495 & 0.048 & 0.048 & 0.019& 0.247 & 1\\
  & & & RSRC & -7.617 & 0.058 & 0.059 & 0.006& 0.847 & 1\\
  & & & GRRC & 0.087 & 0.074 & 0.073 & 0.005& 0.957 & 1\\
  & & & GRN &-0.029 & 0.074 & 0.072 & 0.005& 0.957 & 1\\
  & & & Naive &2.741 & 0.032 & 0.033 & 0.002& 0.902 & 1\\
  & & & Complete &1.385 & 0.104 & 0.105 & 0.011& 0.954 & 1\\
  \\
  & log(3) & & True & 0.043 & 0.033 & 0.033 & 0.001 & 0.949 & 1\\
  \\
  & & 0.5 & RC & -26.719 & 0.048 & 0.048 & 0.037& 0.030 & 1\\
  & & & RSRC & -18.712 & 0.055 & 0.057 & 0.020& 0.343 & 1\\
  & & & GRRC & -0.851 & 0.086 & 0.087 & 0.008& 0.944 & 1\\
  & & & GRN &-1.010 & 0.084 & 0.082 & 0.007& 0.948 & 1\\
  & & & Naive &0.144 & 0.032 & 0.037 & 0.001& 0.913 & 1\\
  & & & Complete &1.284 & 0.106 & 0.108 & 0.012& 0.952 & 1\\
  & & 1 & RC & -32.951 & 0.051 & 0.051 & 0.055& 0.006 & 1\\
  & & & RSRC & -22.881 & 0.060 & 0.062 & 0.029& 0.264 & 1\\
  & & & GRRC & -0.793 & 0.090 & 0.088 & 0.008& 0.946 & 1\\
  & & & GRN &-0.866 & 0.089 & 0.088 & 0.008& 0.947 & 1\\
  & & & Naive &-10.777 & 0.032 & 0.036 & 0.007& 0.362 & 1\\
  & & & Complete &1.298 & 0.106 & 0.108 & 0.012& 0.955 & 1\\
  \\
  75 & log(1.5) & & True & 0.130 & 0.056 & 0.056 & 0.003 & 0.954 & 1\\
  \\
  & & 0.5 & RC & -14.874 & 0.079 & 0.079 & 0.017& 0.72 & 1\\
  & & & RSRC & -12.248 & 0.087 & 0.090 & 0.015& 0.816 & 1\\
  & & & GRRC & -0.101 & 0.121 & 0.119 & 0.014& 0.954 & 0.998\\
  & & & GRN &-0.707 & 0.129 & 0.128 & 0.016& 0.952 & 0.998\\
  & & & Naive &32.244 & 0.057 & 0.059 & 0.053& 0.020 & 1\\
  & & & Complete &1.962 & 0.182 & 0.190 & 0.036& 0.944 & 0.978\\
  & & 1 & RC & -17.226 & 0.082 & 0.082 & 0.021& 0.681 & 1\\
  & & & RSRC & -14.946 & 0.090 & 0.094 & 0.020& 0.782 & 1\\
  & & & GRRC & -0.390 & 0.127 & 0.124 & 0.015& 0.954 & 0.998\\
  & & & GRN &-1.010 & 0.131 & 0.13 & 0.017& 0.946 & 0.998\\
  & & & Naive &17.760 & 0.056 & 0.058 & 0.019& 0.400 & 1\\
  & & & Complete &1.818 & 0.182 & 0.190 & 0.036& 0.944 & 0.975\\
  \\
  & log(3) & & True & 0.188 & 0.054 & 0.055 & 0.003 & 0.948 & 1\\
  \\
  & & 0.5 & RC & -30.268 & 0.083 & 0.084 & 0.051& 0.288 & 1\\
  & & & RSRC & -27.685 & 0.092 & 0.096 & 0.046& 0.443 & 1\\
  & & & GRRC & -1.068 & 0.148 & 0.145 & 0.021& 0.944 & 0.992\\
  & & & GRN &-1.746 & 0.152 & 0.149 & 0.022& 0.944 & 0.992\\
  & & & Naive &20.111 & 0.055 & 0.062 & 0.023& 0.297 & 1\\
  & & & Complete &2.265 & 0.178 & 0.186 & 0.035& 0.948 & 0.976\\
  & & 1 & RC & -32.691 & 0.085 & 0.087 & 0.059& 0.237 & 0.999\\
  & & & RSRC & -30.628 & 0.094 & 0.099 & 0.055& 0.383 & 1\\
  & & & GRRC & -1.096 & 0.152 & 0.150 & 0.022& 0.950 & 0.99\\
  & & & GRN &-1.890 & 0.154 & 0.153 & 0.024& 0.943 & 0.99\\
  & & & Naive &3.982 & 0.054 & 0.061 & 0.004& 0.880 & 1\\
  & & & Complete &2.121 & 0.180 & 0.188 & 0.036& 0.944 & 0.978\\
  \end{tabular}
}
\end{table}

\begin{table}[H]
\renewcommand{\thetable}{\arabic{table}}
\caption{Simulation results for $\beta_Z$ under additive, general measurement error in the outcome and covariate $X$ with $\beta_X = \log 1.5$, normally distributed error, and 25$\%$ censoring for the true event time. For $2000$ simulated data sets, the bias, average bootstrap standard error (ASE) for the 4 proposed estimators, average model standard error (ASE) for naive and complete case, empirical standard error (ESE), mean squared error (MSE), and 95$\%$ coverage probabilities (CP) are presented.}
\label{correlated_err_table1_Z}
\centering
\scalebox{0.71}{
  \begin{tabular}{ccccccccccH}
  
  $\beta_X$ & $\sigma^{2}_{\nu}$ & $\sigma^2_{\epsilon}$ & $\sigma_{\nu,\epsilon}$ & Method & $\%$ Bias & ASE & ESE & MSE & CP & Power \\
  \hline
  log(1.5) & & & & True & 0.072 & 0.032 & 0.033 & 0.001 & 0.949 & 1\\
  \\
  & 0.5 & 0.5 & 0.15 & RC & -13.936 & 0.050 & 0.049 & 0.012& 0.510 & 1\\
  & & & & RSRC & -5.554 & 0.058 & 0.058 & 0.005& 0.901 & 1\\
  & & & & GRRC & 0.245 & 0.071 & 0.069 & 0.005& 0.964 & 1\\
  & & & & GRN &0.115 & 0.070 & 0.068 & 0.005& 0.963 & 1\\
  & & & & Naive &25.189 & 0.031 & 0.032 & 0.032& 0.000 & 1\\
  & & & & Complete &1.428 & 0.104 & 0.105 & 0.011& 0.956 & 1\\
  \\
  & & & 0.30 & RC & -13.893 & 0.050 & 0.050 & 0.012& 0.526 & 1\\
  & & & & RSRC & -5.324 & 0.058 & 0.059 & 0.005& 0.903 & 1\\
  & & & & GRRC & 0.245 & 0.069 & 0.067 & 0.004& 0.964 & 1\\
  & & & & GRN & 0.058 & 0.069 & 0.066 & 0.004& 0.968 & 1\\
  & & & & Naive & 27.656 & 0.031 & 0.033 & 0.038& 0.000 & 1\\
  & & & & Complete &1.486 & 0.104 & 0.106 & 0.011& 0.951 & 1\\
  \\
  & & 1 & 0.15 & RC & -14.153 & 0.054 & 0.053 & 0.012& 0.568 & 1\\
  & & & & RSRC & -5.713 & 0.062 & 0.062 & 0.006& 0.912 & 1\\
  & & & & GRRC & 0.346 & 0.073 & 0.071 & 0.005& 0.964 & 1\\
  & & & & GRN & 0.245 & 0.072 & 0.070 & 0.005& 0.965 & 1\\
  & & & & Naive & 26.113 & 0.031 & 0.032 & 0.034& 0.000 & 1\\
  & & & & Complete &1.457 & 0.104 & 0.106 & 0.011& 0.953 & 1\\
  \\
  & & & 0.30 & RC & -14.138 & 0.055 & 0.053 & 0.012& 0.578 & 1\\
  & & & & RSRC & -5.526 & 0.062 & 0.062 & 0.005& 0.917 & 1\\
  & & & & GRRC & 0.332 & 0.072 & 0.069 & 0.005& 0.966 & 1\\
  & & & & GRN & 0.216 & 0.071 & 0.068 & 0.005& 0.966 & 1\\
  & & & & Naive & 27.786 & 0.031 & 0.032 & 0.038& 0.000 & 1\\
  & & & & Complete &1.428 & 0.104 & 0.106 & 0.011& 0.954 & 1\\
  \\
  & 1 & 0.5 & 0.15 & RC & -19.563 & 0.054 & 0.053 & 0.021& 0.288 & 1\\
  & & & & RSRC & -8.094 & 0.065 & 0.066 & 0.008& 0.850 & 1\\
  & & & & GRRC & 0.231 & 0.078 & 0.076 & 0.006& 0.962 & 1\\
  & & & & GRN &0.115 & 0.077 & 0.075 & 0.006& 0.960 & 1\\
  & & & & Naive &15.581 & 0.030 & 0.031 & 0.013& 0.047 & 1\\
  & & & & Complete &1.385 & 0.104 & 0.105 & 0.011& 0.954 & 1\\
  \\
  & & & 0.30 & RC & -19.606 & 0.055 & 0.054 & 0.021& 0.300 & 1\\
  & & & & RSRC & -7.906 & 0.065 & 0.066 & 0.007& 0.867 & 1\\
  & & & & GRRC & 0.216 & 0.077 & 0.075 & 0.006& 0.958 & 1\\
  & & & & GRN & 0.058 & 0.077 & 0.074 & 0.006& 0.964 & 1\\
  & & & & Naive & 17.731 & 0.030 & 0.031 & 0.016& 0.020 & 1\\
  & & & & Complete &1.443 & 0.104 & 0.106 & 0.011& 0.954 & 1\\
  \\
  & & 1 & 0.15 & RC & -19.707 & 0.059 & 0.058 & 0.022& 0.360 & 1\\
  & & & & RSRC & -8.094 & 0.070 & 0.070 & 0.008& 0.881 & 1\\
  & & & & GRRC & 0.317 & 0.080 & 0.078 & 0.006& 0.960 & 1\\
  & & & & GRN & 0.202 & 0.079 & 0.077 & 0.006& 0.960 & 1\\
  & & & & Naive & 16.504 & 0.030 & 0.031 & 0.014& 0.030 & 1\\
  & & & & Complete &1.472 & 0.104 & 0.106 & 0.011& 0.955 & 1\\
  \\
  & & & 0.30 & RC & -19.736 & 0.060 & 0.058 & 0.022& 0.359 & 1\\
  & & & & RSRC & -7.920 & 0.070 & 0.070 & 0.008& 0.886 & 1\\
  & & & & GRRC & 0.260 & 0.079 & 0.077 & 0.006& 0.961 & 1\\
  & & & & GRN & 0.159 & 0.078 & 0.076 & 0.006& 0.962 & 1\\
  & & & & Naive & 17.99 & 0.030 & 0.031 & 0.016& 0.014 & 1\\
  & & & & Complete &1.371 & 0.104 & 0.106 & 0.011& 0.956 & 1\\
  \end{tabular}
}
\end{table}

\begin{table}[H]
\renewcommand{\thetable}{\arabic{table}}
\caption{Simulation results for $\beta_Z$ under correlated, additive measurement error in the outcome and covariate $X$ with $\beta_X = \log 1.5$, normally distributed error, and 75$\%$ censoring for the true event time. For $2000$ simulated data sets, the bias, average bootstrap standard error (ASE) for the 4 proposed estimators, average model standard error (ASE) for naive and complete case, empirical standard error (ESE), mean squared error (MSE), and 95$\%$ coverage probabilities (CP) are presented.}
\label{correlated_err_table2_Z}
\centering
\scalebox{0.71}{
  \begin{tabular}{ccccccccccH}
  
  $\beta_X$ & $\sigma^{2}_{\nu}$ & $\sigma^2_{\epsilon}$ & $\sigma_{\nu,\epsilon}$ & Method & $\%$ Bias & ASE & ESE & MSE & CP & Power \\
  \hline
  log(1.5) & & & & True & 0.173 & 0.056 & 0.056 & 0.003 & 0.954 & 1\\
  \\
  & 0.5 & 0.5 & 0.15 & RC & -15.105 & 0.088 & 0.087 & 0.019& 0.785 & 1\\
  & & & & RSRC & -12.220 & 0.098 & 0.099 & 0.017& 0.853 & 1\\
  & & & & GRRC & -0.014 & 0.131 & 0.130 & 0.017& 0.954 & 0.998\\
  & & & & GRN &-0.692 & 0.139 & 0.139 & 0.019& 0.950 & 0.996\\
  & & & & Naive &45.272 & 0.051 & 0.053 & 0.101& 0.000 & 1\\
  & & & & Complete &1.962 & 0.182 & 0.190 & 0.036& 0.944 & 0.978\\
  \\
  & & & 0.30 & RC & -14.917 & 0.088 & 0.087 & 0.018& 0.802 & 1\\
  & & & & RSRC & -11.902 & 0.098 & 0.100 & 0.017& 0.858 & 1\\
  & & & & GRRC & 0.043 & 0.129 & 0.128 & 0.016& 0.956 & 0.999\\
  & & & & GRN & -0.808 & 0.137 & 0.137 & 0.019& 0.946 & 0.998\\
  & & & & Naive & 55.457 & 0.052 & 0.055 & 0.151& 0.000 & 1\\
  & & & & Complete &1.861 & 0.182 & 0.191 & 0.037& 0.940 & 0.97\\
  \\
  & & 1 & 0.15 & RC & -15.163 & 0.097 & 0.095 & 0.020& 0.818 & 0.999\\
  & & & & RSRC & -12.119 & 0.107 & 0.108 & 0.019& 0.887 & 1\\
  & & & & GRRC & -0.014 & 0.136 & 0.135 & 0.018& 0.954 & 0.998\\
  & & & & GRN & -0.548 & 0.143 & 0.143 & 0.020& 0.950 & 0.996\\
  & & & & Naive & 44.103 & 0.051 & 0.053 & 0.096& 0.000 & 1\\
  & & & & Complete &2.265 & 0.182 & 0.191 & 0.037& 0.948 & 0.98\\
  \\
  & & & 0.30 & RC & -14.975 & 0.097 & 0.094 & 0.020& 0.827 & 0.999\\
  & & & & RSRC & -12.047 & 0.107 & 0.108 & 0.018& 0.883 & 0.998\\
  & & & & GRRC & -0.144 & 0.135 & 0.132 & 0.018& 0.956 & 0.997\\
  & & & & GRN & -0.779 & 0.142 & 0.141 & 0.020& 0.950 & 0.995\\
  & & & & Naive & 50.595 & 0.051 & 0.053 & 0.126& 0.000 & 1\\
  & & & & Complete &2.049 & 0.182 & 0.187 & 0.035& 0.950 & 0.979\\
  \\
  & 1 & 0.5 & 0.15 & RC & -17.543 & 0.092 & 0.091 & 0.023& 0.742 & 1\\
  & & & & RSRC & -15.018 & 0.101 & 0.104 & 0.022& 0.813 & 1\\
  & & & & GRRC & -0.274 & 0.137 & 0.134 & 0.018& 0.955 & 0.996\\
  & & & & GRN &-0.923 & 0.141 & 0.140 & 0.020& 0.950 & 0.997\\
  & & & & Naive &30.701 & 0.050 & 0.052 & 0.048& 0.010 & 1\\
  & & & & Complete &1.818 & 0.182 & 0.190 & 0.036& 0.944 & 0.975\\
  \\
  & & & 0.30 & RC & -17.586 & 0.092 & 0.090 & 0.023& 0.748 & 1\\
  & & & & RSRC & -14.903 & 0.101 & 0.104 & 0.021& 0.830 & 0.999\\
  & & & & GRRC & -0.115 & 0.134 & 0.133 & 0.018& 0.954 & 0.998\\
  & & & & GRN & -0.822 & 0.140 & 0.138 & 0.019& 0.948 & 0.996\\
  & & & & Naive & 35.894 & 0.050 & 0.052 & 0.065& 0.000 & 1\\
  & & & & Complete &1.847 & 0.181 & 0.185 & 0.034& 0.944 & 0.976\\
  \\
  & & 1 & 0.15 & RC & -17.644 & 0.100 & 0.098 & 0.025& 0.780 & 0.999\\
  & & & & RSRC & -14.816 & 0.111 & 0.113 & 0.023& 0.846 & 0.997\\
  & & & & GRRC & -0.188 & 0.139 & 0.140 & 0.020& 0.944 & 0.996\\
  & & & & GRN & -0.649 & 0.144 & 0.145 & 0.021& 0.944 & 0.996\\
  & & & & Naive & 30.441 & 0.049 & 0.052 & 0.047& 0.010 & 1\\
  & & & & Complete &1.760 & 0.181 & 0.190 & 0.036& 0.941 & 0.971\\
  \\
  & & & 0.30 & RC & -17.500 & 0.101 & 0.098 & 0.024& 0.789 & 0.998\\
  & & & & RSRC & -14.946 & 0.110 & 0.113 & 0.024& 0.849 & 0.994\\
  & & & & GRRC & -0.144 & 0.140 & 0.138 & 0.019& 0.955 & 0.996\\
  & & & & GRN & -0.750 & 0.144 & 0.145 & 0.021& 0.946 & 0.996\\
  & & & & Naive & 34.192 & 0.050 & 0.052 & 0.059& 0.001 & 1\\
  & & & & Complete &1.746 & 0.182 & 0.190 & 0.036& 0.946 & 0.974\\
  \end{tabular}
}
\end{table}

\begin{table}[H]
\renewcommand{\thetable}{\arabic{table}}
\caption{Simulation results for $\beta_Z$ under correlated, additive measurement error in the outcome and covariate $X$ with $\beta_X=\log 3$, normally distributed error, and 25$\%$ censoring for the true event time. For $2000$ simulated data sets, the bias, average bootstrap standard error (ASE) for the 4 proposed estimators, average model standard error (ASE) for naive and complete case, empirical standard error (ESE), mean squared error (MSE), and 95$\%$ coverage probabilities (CP) are presented.}
\label{correlated_err_table3_Z}
\centering
\scalebox{0.71}{
  \begin{tabular}{ccccccccccH}
  
  $\beta_X$ & $\sigma^{2}_{\nu}$ & $\sigma^2_{\epsilon}$ & $\sigma_{\nu,\epsilon}$ & Method & $\%$ Bias & ASE & ESE & MSE & CP & Power \\
  \hline
  log(3) & & & & True & 0.043 & 0.033 & 0.033 & 0.001 & 0.945 & 1\\
  \\
  & 0.5 & 0.5 & 0.15 & RC & -31.667 & 0.060 & 0.060 & 0.052& 0.036 & 1\\
  & & & & RSRC & -20.789 & 0.070 & 0.072 & 0.026& 0.466 & 1\\
  & & & & GRRC & -0.433 & 0.088 & 0.086 & 0.007& 0.952 & 1\\
  & & & & GRN &-0.548 & 0.087 & 0.084 & 0.007& 0.952 & 1\\
  & & & & Naive &26.892 & 0.031 & 0.034 & 0.036& 0.000 & 1\\
  & & & & Complete &1.284 & 0.106 & 0.108 & 0.012& 0.952 & 1\\
  \\
  & & & 0.30 & RC & -32.432 & 0.063 & 0.062 & 0.054& 0.036 & 1\\
  & & & & RSRC & -20.602 & 0.072 & 0.074 & 0.026& 0.497 & 1\\
  & & & & GRRC & -0.303 & 0.087 & 0.084 & 0.007& 0.956 & 1\\
  & & & & GRN & -0.491 & 0.087 & 0.084 & 0.007& 0.957 & 1\\
  & & & & Naive & 26.661 & 0.031 & 0.034 & 0.035& 0.000 & 1\\
  & & & & Complete &1.284 & 0.106 & 0.107 & 0.012& 0.950 & 1\\
  \\
  & & 1 & 0.15 & RC & -32.720 & 0.067 & 0.066 & 0.056& 0.050 & 1\\
  & & & & RSRC & -20.977 & 0.076 & 0.077 & 0.027& 0.522 & 0.999\\
  & & & & GRRC & -0.303 & 0.088 & 0.087 & 0.008& 0.955 & 1\\
  & & & & GRN & -0.404 & 0.087 & 0.085 & 0.007& 0.959 & 1\\
  & & & & Naive & 29.316 & 0.031 & 0.034 & 0.042& 0.000 & 1\\
  & & & & Complete &1.298 & 0.106 & 0.108 & 0.012& 0.953 & 1\\
  \\
  & & & 0.30 & RC & -33.225 & 0.069 & 0.068 & 0.058& 0.050 & 1\\
  & & & & RSRC & -20.789 & 0.077 & 0.078 & 0.027& 0.550 & 0.999\\
  & & & & GRRC & -0.188 & 0.088 & 0.086 & 0.007& 0.956 & 1\\
  & & & & GRN & -0.317 & 0.087 & 0.085 & 0.007& 0.954 & 1\\
  & & & & Naive & 29.128 & 0.031 & 0.034 & 0.042& 0.000 & 1\\
  & & & & Complete &1.313 & 0.106 & 0.108 & 0.012& 0.950 & 1\\
  \\
  & 1 & 0.5 & 0.15 & RC & -36.341 & 0.062 & 0.062 & 0.067& 0.010 & 1\\
  & & & & RSRC & -23.920 & 0.073 & 0.074 & 0.033& 0.371 & 1\\
  & & & & GRRC & -0.375 & 0.092 & 0.090 & 0.008& 0.954 & 1\\
  & & & & GRN &-0.519 & 0.091 & 0.089 & 0.008& 0.955 & 1\\
  & & & & Naive &16.605 & 0.030 & 0.034 & 0.014& 0.040 & 1\\
  & & & & Complete &1.298 & 0.106 & 0.108 & 0.012& 0.955 & 1\\
  \\
  & & & 0.30 & RC & -37.048 & 0.064 & 0.063 & 0.070& 0.008 & 1\\
  & & & & RSRC & -23.761 & 0.074 & 0.076 & 0.033& 0.398 & 1\\
  & & & & GRRC & -0.361 & 0.091 & 0.089 & 0.008& 0.956 & 0.999\\
  & & & & GRN & -0.447 & 0.091 & 0.089 & 0.008& 0.953 & 1\\
  & & & & Naive & 16.764 & 0.030 & 0.033 & 0.015& 0.038 & 1\\
  & & & & Complete &1.356 & 0.106 & 0.108 & 0.012& 0.948 & 1\\
  \\
  & & 1 & 0.15 & RC & -37.063 & 0.069 & 0.068 & 0.071& 0.018 & 1\\
  & & & & RSRC & -23.660 & 0.079 & 0.080 & 0.033& 0.454 & 0.998\\
  & & & & GRRC & -0.274 & 0.092 & 0.090 & 0.008& 0.955 & 1\\
  & & & & GRN & -0.361 & 0.091 & 0.089 & 0.008& 0.952 & 1\\
  & & & & Naive & 19.274 & 0.030 & 0.033 & 0.019& 0.012 & 1\\
  & & & & Complete &1.284 & 0.106 & 0.108 & 0.012& 0.950 & 1\\
  \\
  & & & 0.30 & RC & -37.524 & 0.070 & 0.069 & 0.072& 0.017 & 0.998\\
  & & & & RSRC & -23.574 & 0.080 & 0.081 & 0.033& 0.467 & 0.998\\
  & & & & GRRC & -0.202 & 0.092 & 0.090 & 0.008& 0.956 & 1\\
  & & & & GRN & -0.289 & 0.091 & 0.089 & 0.008& 0.956 & 1\\
  & & & & Naive & 19.361 & 0.030 & 0.033 & 0.019& 0.012 & 1\\
  & & & & Complete &1.327 & 0.106 & 0.108 & 0.012& 0.950 & 1\\
  \\
  \end{tabular}
}
\end{table}

\section*{Appendix D: Classical measurement error tables}

\begin{table}[H]
\renewcommand{\thetable}{\arabic{table}}
\caption{Simulation results for $\beta_X=\log 1.5$ under correlated additive measurement error in the outcome and classical measurement error in the covariate $X$ with normally distributed error and 25$\%$ censoring for the true event time. For $2000$ simulated data sets, the bias, average bootstrap standard error (ASE) for the 4 proposed estimators, average model standard error (ASE) for naive and complete case, empirical standard error (ESE), mean squared error (MSE), and 95$\%$ coverage probabilities (CP) are presented.}
\centering
\scalebox{0.68}{
  \begin{tabular}{ccccccccccH}
  
  $\beta_X$ & $\sigma^{2}_{\nu}$ & $\sigma^2_{\epsilon}$ & $\sigma_{\nu,\epsilon}$ & Method & $\%$ Bias & ASE & ESE & MSE & CP & Power \\
  \hline
  log(1.5) & & & & True & -0.049 & 0.030 & 0.031 & 0.001 & 0.946 & 1\\
  \\
  & 0.5 & 0.5 & 0.15 & RC & -13.762 & 0.056 & 0.057 & 0.006& 0.800 & 1\\
  & & & & RSRC & -6.141 & 0.066 & 0.065 & 0.005& 0.920 & 1\\
  & & & & GRRC & 0.123 & 0.081 & 0.082 & 0.007& 0.949 & 0.998\\
  & & & & GRN &0.296 & 0.081 & 0.082 & 0.007& 0.947 & 0.998\\
  & & & & Naive &-78.428 & 0.023 & 0.023 & 0.102& 0.000 & 0.964\\
  & & & & Complete &0.123 & 0.098 & 0.099 & 0.010& 0.952 & 0.986\\
  \\
  & & & 0.30 & RC & -13.589 & 0.057 & 0.057 & 0.006& 0.800 & 1\\
  & & & & RSRC & -5.944 & 0.068 & 0.066 & 0.005& 0.929 & 1\\
  & & & & GRRC & 0.222 & 0.081 & 0.082 & 0.007& 0.945 & 0.998\\
  & & & & GRN & 0.518 & 0.081 & 0.082 & 0.007& 0.942 & 0.999\\
  & & & & Naive & -93.621 & 0.023 & 0.024 & 0.145& 0.000 & 0.215\\
  & & & & Complete &0.148 & 0.098 & 0.099 & 0.010& 0.954 & 0.988\\
  \\
  & & 1 & 0.15 & RC & -13.836 & 0.067 & 0.067 & 0.008& 0.832 & 1\\
  & & & & RSRC & -6.758 & 0.079 & 0.078 & 0.007& 0.918 & 0.998\\
  & & & & GRRC & 0.000 & 0.087 & 0.088 & 0.008& 0.946 & 0.995\\
  & & & & GRN & 0.148 & 0.087 & 0.088 & 0.008& 0.946 & 0.996\\
  & & & & Naive & -84.594 & 0.019 & 0.020 & 0.118& 0.000 & 0.896\\
  & & & & Complete &0.173 & 0.098 & 0.099 & 0.010& 0.952 & 0.988\\
  \\
  & & & 0.30 & RC & -13.688 & 0.068 & 0.068 & 0.008& 0.836 & 1\\
  & & & & RSRC & -6.708 & 0.080 & 0.079 & 0.007& 0.912 & 1\\
  & & & & GRRC & 0.247 & 0.087 & 0.088 & 0.008& 0.948 & 0.996\\
  & & & & GRN & 0.469 & 0.087 & 0.088 & 0.008& 0.944 & 0.997\\
  & & & & Naive & -95.471 & 0.019 & 0.020 & 0.150& 0.000 & 0.161\\
  & & & & Complete &0.271 & 0.098 & 0.098 & 0.010& 0.952 & 0.988\\
  \\
  & 1 & 0.5 & 0.15 & RC & -19.286 & 0.062 & 0.062 & 0.010& 0.734 & 1\\
  & & & & RSRC & -9.224 & 0.074 & 0.073 & 0.007& 0.907 & 0.997\\
  & & & & GRRC & 0.148 & 0.083 & 0.084 & 0.007& 0.942 & 0.997\\
  & & & & GRN &0.271 & 0.083 & 0.084 & 0.007& 0.943 & 0.997\\
  & & & & Naive &-78.552 & 0.023 & 0.023 & 0.102& 0.000 & 0.962\\
  & & & & Complete &0.247 & 0.098 & 0.098 & 0.010& 0.954 & 0.987\\
  \\
  & & & 0.30 & RC & -19.286 & 0.062 & 0.063 & 0.010& 0.732 & 1\\
  & & & & RSRC & -9.372 & 0.076 & 0.073 & 0.007& 0.904 & 0.999\\
  & & & & GRRC & 0.197 & 0.083 & 0.084 & 0.007& 0.944 & 0.998\\
  & & & & GRN & 0.370 & 0.083 & 0.084 & 0.007& 0.946 & 0.998\\
  & & & & Naive & -91.993 & 0.023 & 0.023 & 0.140& 0.000 & 0.306\\
  & & & & Complete &0.173 & 0.098 & 0.099 & 0.010& 0.948 & 0.988\\
  \\
  & & 1 & 0.15 & RC & -19.139 & 0.074 & 0.074 & 0.012& 0.791 & 0.996\\
  & & & & RSRC & -10.013 & 0.088 & 0.087 & 0.009& 0.907 & 0.99\\
  & & & & GRRC & 0.025 & 0.089 & 0.090 & 0.008& 0.942 & 0.996\\
  & & & & GRN & 0.123 & 0.089 & 0.090 & 0.008& 0.944 & 0.996\\
  & & & & Naive & -84.619 & 0.019 & 0.020 & 0.118& 0.000 & 0.894\\
  & & & & Complete &0.271 & 0.098 & 0.099 & 0.010& 0.954 & 0.986\\
  \\
  & & & 0.30 & RC & -19.262 & 0.074 & 0.074 & 0.012& 0.779 & 0.997\\
  & & & & RSRC & -10.137 & 0.089 & 0.088 & 0.009& 0.902 & 0.989\\
  & & & & GRRC & 0.099 & 0.089 & 0.090 & 0.008& 0.944 & 0.996\\
  & & & & GRN & 0.247 & 0.089 & 0.090 & 0.008& 0.943 & 0.994\\
  & & & & Naive & -94.336 & 0.019 & 0.020 & 0.147& 0.000 & 0.226\\
  & & & & Complete &0.247 & 0.098 & 0.098 & 0.010& 0.953 & 0.987\\
  \end{tabular}
}
\end{table}

\begin{table}[H]
\renewcommand{\thetable}{\arabic{table}}
\caption{Simulation results for $\beta_X=\log 1.5$ under correlated additive measurement error in the outcome and classical measurement error in the covariate $X$ with normally distributed error and 75$\%$ censoring for the true event time. For $2000$ simulated data sets, the bias, average bootstrap standard error (ASE) for the 4 proposed estimators, average model standard error (ASE) for naive and complete case, empirical standard error (ESE), mean squared error (MSE), and 95$\%$ coverage probabilities (CP) are presented.}
\centering
\scalebox{0.71}{
  \begin{tabular}{ccccccccccH}
  
  $\beta_X$ & $\sigma^{2}_{\nu}$ & $\sigma^2_{\epsilon}$ & $\sigma_{\nu,\epsilon}$ & Method & $\%$ Bias & ASE & ESE & MSE & CP & Power \\
  \hline
  log(1.5) & & & & True & 0.123 & 0.054 & 0.054 & 0.003 & 0.949 & 1\\
  \\
  & 0.5 & 0.5 & 0.15 & RC & -14.946 & 0.104 & 0.103 & 0.014& 0.901 & 0.917\\
  & & & & RSRC & -12.406 & 0.114 & 0.113 & 0.015& 0.918 & 0.886\\
  & & & & GRRC & 0.148 & 0.151 & 0.149 & 0.022& 0.956 & 0.783\\
  & & & & GRN &0.789 & 0.153 & 0.152 & 0.023& 0.952 & 0.776\\
  & & & & Naive &-115.916 & 0.043 & 0.043 & 0.223& 0.000 & 0.332\\
  & & & & Complete &0.543 & 0.177 & 0.183 & 0.033& 0.951 & 0.635\\
  \\
  & & & 0.30 & RC & -14.675 & 0.104 & 0.103 & 0.014& 0.896 & 0.934\\
  & & & & RSRC & -12.011 & 0.115 & 0.112 & 0.015& 0.925 & 0.891\\
  & & & & GRRC & -0.296 & 0.150 & 0.147 & 0.022& 0.956 & 0.789\\
  & & & & GRN & 1.233 & 0.149 & 0.147 & 0.022& 0.953 & 0.808\\
  & & & & Naive & -156.462 & 0.045 & 0.045 & 0.404& 0.000 & 0.999\\
  & & & & Complete &0.123 & 0.176 & 0.181 & 0.033& 0.952 & 0.629\\
  \\
  & & 1 & 0.15 & RC & -14.970 & 0.124 & 0.123 & 0.019& 0.918 & 0.804\\
  & & & & RSRC & -12.677 & 0.137 & 0.138 & 0.022& 0.926 & 0.748\\
  & & & & GRRC & -0.370 & 0.162 & 0.160 & 0.026& 0.954 & 0.723\\
  & & & & GRN & 0.074 & 0.164 & 0.163 & 0.026& 0.956 & 0.706\\
  & & & & Naive & -111.082 & 0.036 & 0.036 & 0.204& 0.000 & 0.244\\
  & & & & Complete &-0.074 & 0.176 & 0.180 & 0.032& 0.947 & 0.626\\
  \\
  & & & 0.30 & RC & -14.477 & 0.124 & 0.123 & 0.019& 0.919 & 0.816\\
  & & & & RSRC & -12.529 & 0.137 & 0.137 & 0.021& 0.929 & 0.76\\
  & & & & GRRC & -0.074 & 0.162 & 0.160 & 0.026& 0.958 & 0.716\\
  & & & & GRN & 0.715 & 0.164 & 0.162 & 0.026& 0.955 & 0.706\\
  & & & & Naive & -138.212 & 0.037 & 0.038 & 0.315& 0.000 & 0.985\\
  & & & & Complete &0.247 & 0.176 & 0.181 & 0.033& 0.948 & 0.626\\
  \\
  & 1 & 0.5 & 0.15 & RC & -17.091 & 0.108 & 0.107 & 0.016& 0.896 & 0.89\\
  & & & & RSRC & -15.587 & 0.117 & 0.118 & 0.018& 0.901 & 0.837\\
  & & & & GRRC & -0.074 & 0.153 & 0.151 & 0.023& 0.960 & 0.77\\
  & & & & GRN &0.666 & 0.154 & 0.152 & 0.023& 0.956 & 0.776\\
  & & & & Naive &-99.367 & 0.042 & 0.042 & 0.164& 0.000 & 0.056\\
  & & & & Complete &0.617 & 0.177 & 0.181 & 0.033& 0.952 & 0.627\\
  \\
  & & & 0.30 & RC & -17.042 & 0.108 & 0.108 & 0.016& 0.890 & 0.893\\
  & & & & RSRC & -15.538 & 0.118 & 0.119 & 0.018& 0.901 & 0.845\\
  & & & & GRRC & -0.173 & 0.153 & 0.151 & 0.023& 0.956 & 0.766\\
  & & & & GRN & 0.987 & 0.154 & 0.152 & 0.023& 0.950 & 0.774\\
  & & & & Naive & -126.003 & 0.044 & 0.044 & 0.263& 0.000 & 0.672\\
  & & & & Complete &0.592 & 0.178 & 0.183 & 0.034& 0.950 & 0.623\\
  \\
  & & 1 & 0.15 & RC & -16.993 & 0.129 & 0.127 & 0.021& 0.910 & 0.758\\
  & & & & RSRC & -15.784 & 0.140 & 0.142 & 0.024& 0.910 & 0.687\\
  & & & & GRRC & 0.247 & 0.165 & 0.164 & 0.027& 0.959 & 0.704\\
  & & & & GRN & 0.765 & 0.166 & 0.166 & 0.028& 0.956 & 0.7\\
  & & & & Naive & -99.614 & 0.036 & 0.036 & 0.164& 0.000 & 0.05\\
  & & & & Complete &0.518 & 0.178 & 0.183 & 0.034& 0.947 & 0.628\\
  \\
  & & & 0.30 & RC & -17.067 & 0.129 & 0.127 & 0.021& 0.908 & 0.756\\
  & & & & RSRC & -15.316 & 0.141 & 0.141 & 0.024& 0.914 & 0.698\\
  & & & & GRRC & -0.222 & 0.164 & 0.162 & 0.026& 0.962 & 0.706\\
  & & & & GRN & 0.567 & 0.165 & 0.164 & 0.027& 0.963 & 0.713\\
  & & & & Naive & -118.136 & 0.036 & 0.037 & 0.231& 0.000 & 0.522\\
  & & & & Complete &0.222 & 0.177 & 0.182 & 0.033& 0.947 & 0.63\\
  \end{tabular}
}
\end{table}

\begin{table}[H]
\renewcommand{\thetable}{\arabic{table}}
\caption{Simulation results for $\beta_X=\log 3$ under correlated additive measurement error in the outcome and classical measurement error in the covariate $X$ with normally distributed error and 25$\%$ censoring for the true event time. For $2000$ simulated data sets, the bias, average bootstrap standard error (ASE) for the 4 proposed estimators, average model standard error (ASE) for naive and complete case, empirical standard error (ESE), mean squared error (MSE), and 95$\%$ coverage probabilities (CP) are presented.}
\centering
\scalebox{0.71}{
  \begin{tabular}{ccccccccccH}
  
  $\beta_X$ & $\sigma^{2}_{\nu}$ & $\sigma^2_{\epsilon}$ & $\sigma_{\nu,\epsilon}$ & Method & $\%$ Bias & ASE & ESE & MSE & CP & Power \\
  \hline
  log(3) & & & & True & 0.046 & 0.037 & 0.036 & 0.001 & 0.955 & 1\\
  \\
  & 0.5 & 0.5 & 0.15 & RC & -30.921 & 0.073 & 0.074 & 0.121& 0.018 & 1\\
  & & & & RSRC & -22.665 & 0.088 & 0.087 & 0.070& 0.216 & 1\\
  & & & & GRRC & 0.200 & 0.112 & 0.110 & 0.012& 0.952 & 1\\
  & & & & GRN &0.291 & 0.112 & 0.111 & 0.012& 0.954 & 1\\
  & & & & Naive &-70.016 & 0.023 & 0.026 & 0.592& 0.000 & 1\\
  & & & & Complete &0.792 & 0.118 & 0.118 & 0.014& 0.957 & 1\\
  \\
  & & & 0.30 & RC & -31.658 & 0.076 & 0.076 & 0.127& 0.024 & 1\\
  & & & & RSRC & -22.774 & 0.092 & 0.091 & 0.071& 0.240 & 1\\
  & & & & GRRC & 0.300 & 0.112 & 0.110 & 0.012& 0.957 & 1\\
  & & & & GRN & 0.319 & 0.112 & 0.110 & 0.012& 0.954 & 1\\
  & & & & Naive & -76.032 & 0.023 & 0.025 & 0.698& 0.000 & 1\\
  & & & & Complete &0.737 & 0.118 & 0.117 & 0.014& 0.955 & 1\\
  \\
  & & 1 & 0.15 & RC & -31.604 & 0.087 & 0.088 & 0.128& 0.062 & 1\\
  & & & & RSRC & -23.903 & 0.104 & 0.103 & 0.080& 0.304 & 1\\
  & & & & GRRC & 0.401 & 0.115 & 0.113 & 0.013& 0.953 & 1\\
  & & & & GRN & 0.428 & 0.115 & 0.113 & 0.013& 0.952 & 1\\
  & & & & Naive & -78.572 & 0.020 & 0.021 & 0.746& 0.000 & 1\\
  & & & & Complete &0.792 & 0.118 & 0.118 & 0.014& 0.960 & 1\\
  \\
  & & & 0.30 & RC & -32.159 & 0.090 & 0.089 & 0.133& 0.065 & 1\\
  & & & & RSRC & -24.058 & 0.108 & 0.107 & 0.081& 0.328 & 1\\
  & & & & GRRC & 0.437 & 0.115 & 0.113 & 0.013& 0.956 & 1\\
  & & & & GRN & 0.519 & 0.115 & 0.113 & 0.013& 0.952 & 1\\
  & & & & Naive & -82.713 & 0.019 & 0.021 & 0.826& 0.000 & 1\\
  & & & & Complete &0.801 & 0.118 & 0.118 & 0.014& 0.958 & 1\\
  \\
  & 1 & 0.5 & 0.15 & RC & -35.681 & 0.075 & 0.076 & 0.159& 0.008 & 1\\
  & & & & RSRC & -26.488 & 0.090 & 0.090 & 0.093& 0.150 & 1\\
  & & & & GRRC & 0.191 & 0.113 & 0.112 & 0.012& 0.950 & 1\\
  & & & & GRN &0.218 & 0.113 & 0.112 & 0.012& 0.952 & 1\\
  & & & & Naive &-71.244 & 0.023 & 0.025 & 0.613& 0.000 & 1\\
  & & & & Complete &0.746 & 0.118 & 0.118 & 0.014& 0.956 & 1\\
  \\
  & & & 0.30 & RC & -36.382 & 0.077 & 0.077 & 0.166& 0.009 & 1\\
  & & & & RSRC & -26.961 & 0.093 & 0.092 & 0.096& 0.156 & 1\\
  & & & & GRRC & 0.300 & 0.113 & 0.111 & 0.012& 0.954 & 1\\
  & & & & GRN & 0.300 & 0.113 & 0.111 & 0.012& 0.956 & 1\\
  & & & & Naive & -76.360 & 0.023 & 0.025 & 0.704& 0.000 & 1\\
  & & & & Complete &0.737 & 0.118 & 0.118 & 0.014& 0.956 & 1\\
  \\
  & & 1 & 0.15 & RC & -36.055 & 0.090 & 0.090 & 0.165& 0.034 & 1\\
  & & & & RSRC & -27.835 & 0.107 & 0.106 & 0.105& 0.222 & 1\\
  & & & & GRRC & 0.382 & 0.116 & 0.114 & 0.013& 0.948 & 1\\
  & & & & GRN & 0.428 & 0.115 & 0.114 & 0.013& 0.950 & 1\\
  & & & & Naive & -79.437 & 0.020 & 0.021 & 0.762& 0.000 & 1\\
  & & & & Complete &0.801 & 0.118 & 0.118 & 0.014& 0.957 & 1\\
  \\
  & & & 0.30 & RC & -36.564 & 0.091 & 0.091 & 0.170& 0.039 & 1\\
  & & & & RSRC & -28.190 & 0.110 & 0.108 & 0.108& 0.231 & 1\\
  & & & & GRRC & 0.382 & 0.116 & 0.114 & 0.013& 0.952 & 1\\
  & & & & GRN & 0.437 & 0.115 & 0.114 & 0.013& 0.954 & 1\\
  & & & & Naive & -82.977 & 0.019 & 0.021 & 0.831& 0.000 & 1\\
  & & & & Complete &0.765 & 0.118 & 0.118 & 0.014& 0.957 & 1\\
  \\
  \end{tabular}
}
\end{table}

\section*{Appendix E: Gamma distributed error tables}

\begin{table}[H]
\renewcommand{\thetable}{\arabic{table}}
\caption{Simulation results for $\beta_X$ under additive measurement error only in the outcome with gamma distributed error and 25 and 75$\%$ censoring for the true event time. For $2000$ simulated data sets, the bias, average bootstrap standard error (ASE) for the 4 proposed estimators, average model standard error (ASE) for naive and complete case, empirical standard error (ESE), mean squared error (MSE), and 95$\%$ coverage probabilities (CP) are presented.}
\label{outcome_gamma_err_table}
\centering
\scalebox{0.68}{
  \begin{tabular}{cccccccccH}
  
  $\%$ Censoring & $\beta_X$ & $\sigma^{2}_{\nu}$ & Method & $\%$ Bias & ASE & ESE & MSE & CP & Power\\
  \hline
  25 & log(1.5) & & True & 0.099 & 0.030 & 0.032 & 0.001 & 0.942 & 1\\
  \\
  & & 0.5 & RC & -19.558 & 0.045 & 0.045 & 0.008& 0.574 & 1\\
  & & & RSRC & -4.563 & 0.060 & 0.059 & 0.004& 0.935 & 1\\
  & & & GRRC & -0.567 & 0.067 & 0.067 & 0.004& 0.949 & 1\\
  & & & GRN &-0.567 & 0.066 & 0.067 & 0.004& 0.947 & 1\\
  & & & Naive &-31.371 & 0.030 & 0.032 & 0.017& 0.018 & 1\\
  & & & Complete &0.543 & 0.098 & 0.100 & 0.010& 0.952 & 0.99\\
  & & 1 & RC & -28.905 & 0.052 & 0.052 & 0.016& 0.380 & 1\\
  & & & RSRC & -8.879 & 0.071 & 0.071 & 0.006& 0.918 & 0.998\\
  & & & GRRC & -0.617 & 0.075 & 0.076 & 0.006& 0.950 & 1\\
  & & & GRN &-0.592 & 0.075 & 0.076 & 0.006& 0.945 & 1\\
  & & & Naive &-38.869 & 0.029 & 0.032 & 0.026& 0.001 & 1\\
  & & & Complete &0.617 & 0.098 & 0.100 & 0.010& 0.949 & 0.99\\
  \\
  & log(3) & & True & 0.155 & 0.037 & 0.037 & 0.001 & 0.941 & 1 \\
  \\
  & & 0.5 & RC & -33.733 & 0.055 & 0.056 & 0.140& 0.000 & 1\\
  & & & RSRC & -23.156 & 0.067 & 0.069 & 0.069& 0.041 & 1\\
  & & & GRRC & -1.211 & 0.113 & 0.116 & 0.014& 0.923 & 1\\
  & & & GRN &-1.211 & 0.113 & 0.119 & 0.014& 0.920 & 1\\
  & & & Naive &-38.166 & 0.030 & 0.043 & 0.178& 0.000 & 1\\
  & & & Complete &0.819 & 0.119 & 0.121 & 0.015& 0.948 & 1\\
  & & 1 & RC & -41.334 & 0.058 & 0.059 & 0.210& 0.000 & 1\\
  & & & RSRC & -28.254 & 0.074 & 0.076 & 0.102& 0.019 & 1\\
  & & & GRRC & -0.892 & 0.115 & 0.116 & 0.014& 0.936 & 1\\
  & & & GRN &-0.856 & 0.115 & 0.122 & 0.015& 0.928 & 1\\
  & & & Naive &-44.948 & 0.030 & 0.041 & 0.246& 0.000 & 1\\
  & & & Complete &0.874 & 0.119 & 0.122 & 0.015& 0.946 & 1\\
  \\
  75 & log(1.5) & & True & 0.395 & 0.054 & 0.056 & 0.003 & 0.936 & 1\\
  \\
  & & 0.5 & RC & -19.829 & 0.080 & 0.080 & 0.013& 0.834 & 0.978\\
  & & & RSRC & -9.989 & 0.100 & 0.103 & 0.012& 0.921 & 0.913\\
  & & & GRRC & 0.518 & 0.118 & 0.118 & 0.014& 0.954 & 0.948\\
  & & & GRN &0.543 & 0.116 & 0.116 & 0.014& 0.956 & 0.955\\
  & & & Naive &-40.719 & 0.054 & 0.057 & 0.031& 0.156 & 0.99\\
  & & & Complete &2.318 & 0.177 & 0.180 & 0.032& 0.950 & 0.658\\
  & & 1 & RC & -19.903 & 0.089 & 0.091 & 0.015& 0.854 & 0.938\\
  & & & RSRC & -13.762 & 0.112 & 0.119 & 0.017& 0.906 & 0.839\\
  & & & GRRC & 0.641 & 0.121 & 0.120 & 0.014& 0.958 & 0.934\\
  & & & GRN &0.641 & 0.119 & 0.118 & 0.014& 0.952 & 0.946\\
  & & & Naive &-36.279 & 0.054 & 0.058 & 0.025& 0.242 & 0.996\\
  & & & Complete &2.738 & 0.178 & 0.181 & 0.033& 0.948 & 0.65\\
  \\
  & log(3) & & True & 0.300 & 0.058 & 0.059 & 0.003 & 0.948 & 1 \\
  \\
  & & 0.5 & RC & -33.187 & 0.086 & 0.087 & 0.140& 0.010 & 1\\
  & & & RSRC & -28.527 & 0.107 & 0.110 & 0.110& 0.168 & 1\\
  & & & GRRC & -0.692 & 0.174 & 0.176 & 0.031& 0.937 & 0.997\\
  & & & GRN &-0.546 & 0.173 & 0.180 & 0.032& 0.940 & 0.998\\
  & & & Naive &-40.469 & 0.053 & 0.068 & 0.202& 0.000 & 1\\
  & & & Complete &2.458 & 0.193 & 0.200 & 0.041& 0.946 & 1\\
  & & 1 & RC & -33.824 & 0.097 & 0.100 & 0.148& 0.022 & 1\\
  & & & RSRC & -30.957 & 0.121 & 0.128 & 0.132& 0.201 & 0.999\\
  & & & GRRC & -0.628 & 0.176 & 0.183 & 0.034& 0.938 & 0.996\\
  & & & GRN &-0.528 & 0.174 & 0.183 & 0.034& 0.934 & 0.998\\
  & & & Naive &-39.186 & 0.053 & 0.068 & 0.190& 0.000 & 1\\
  & & & Complete &2.485 & 0.193 & 0.204 & 0.042& 0.944 & 1\\
  \end{tabular}
}
\end{table}

\begin{table}[H]
\renewcommand{\thetable}{\arabic{table}}
\caption{Simulation results for $\beta_X=\log 1.5$ under correlated, additive measurement error in the outcome and covariate $X$ with gamma distributed error and 25$\%$ censoring for the true event time. For $2000$ simulated data sets, the bias, average bootstrap standard error (ASE) for the 4 proposed estimators, average model standard error (ASE) for naive and complete case, empirical standard error (ESE), mean squared error (MSE), and 95$\%$ coverage probabilities (CP) are presented.}
\label{correlated_gamma_err_table1}
\centering
\scalebox{0.71}{
  \begin{tabular}{ccccccccccH}
  
  $\beta_X$ & $\sigma^{2}_{\nu}$ & $\sigma^2_{\epsilon}$ & $\sigma_{\nu,\epsilon}$ & Method & $\%$ Bias & ASE & ESE & MSE & CP & Power \\
  \hline
  log(1.5) & & & & True & 0.099 & 0.030 & 0.032 & 0.001 & 0.942 & 1\\
  \\
  & 0.5 & 0.5 & 0.15 & RC & -23.060 & 0.057 & 0.057 & 0.012& 0.601 & 1\\
  & & & & RSRC & -5.944 & 0.075 & 0.076 & 0.006& 0.928 & 0.999\\
  & & & & GRRC & -0.888 & 0.081 & 0.082 & 0.007& 0.945 & 0.999\\
  & & & & GRN &-0.814 & 0.081 & 0.082 & 0.007& 0.943 & 0.999\\
  & & & & Naive &-56.972 & 0.025 & 0.028 & 0.054& 0.000 & 1\\
  & & & & Complete &0.543 & 0.098 & 0.100 & 0.010& 0.952 & 0.99\\
  \\
  & & & 0.30 & RC & -25.206 & 0.058 & 0.058 & 0.014& 0.547 & 1\\
  & & & & RSRC & -4.760 & 0.077 & 0.079 & 0.007& 0.925 & 1\\
  & & & & GRRC & -1.282 & 0.082 & 0.084 & 0.007& 0.941 & 0.998\\
  & & & & GRN & -1.110 & 0.082 & 0.083 & 0.007& 0.943 & 0.998\\
  & & & & Naive & -62.718 & 0.025 & 0.028 & 0.066& 0.000 & 1\\
  & & & & Complete &0.543 & 0.098 & 0.099 & 0.010& 0.952 & 0.992\\
  \\
  & & 1 & 0.15 & RC & -25.403 & 0.068 & 0.067 & 0.015& 0.607 & 1\\
  & & & & RSRC & -8.903 & 0.086 & 0.087 & 0.009& 0.906 & 0.994\\
  & & & & GRRC & -1.726 & 0.087 & 0.089 & 0.008& 0.938 & 0.998\\
  & & & & GRN & -1.578 & 0.086 & 0.088 & 0.008& 0.942 & 0.997\\
  & & & & Naive & -66.689 & 0.022 & 0.025 & 0.074& 0.000 & 1\\
  & & & & Complete &0.469 & 0.098 & 0.100 & 0.010& 0.952 & 0.991\\
  \\
  & & & 0.30 & RC & -27.499 & 0.068 & 0.068 & 0.017& 0.562 & 0.998\\
  & & & & RSRC & -7.941 & 0.088 & 0.091 & 0.009& 0.901 & 0.995\\
  & & & & GRRC & -1.899 & 0.088 & 0.090 & 0.008& 0.934 & 0.996\\
  & & & & GRN & -1.603 & 0.087 & 0.089 & 0.008& 0.938 & 0.996\\
  & & & & Naive & -71.030 & 0.022 & 0.026 & 0.084& 0.000 & 0.998\\
  & & & & Complete &0.641 & 0.098 & 0.100 & 0.010& 0.946 & 0.989\\
  \\
  & 1 & 0.5 & 0.15 & RC & -31.988 & 0.064 & 0.063 & 0.021& 0.468 & 0.993\\
  & & & & RSRC & -9.323 & 0.087 & 0.090 & 0.009& 0.912 & 0.981\\
  & & & & GRRC & -0.863 & 0.085 & 0.086 & 0.007& 0.950 & 0.998\\
  & & & & GRN &-0.789 & 0.085 & 0.086 & 0.007& 0.949 & 0.998\\
  & & & & Naive &-61.189 & 0.025 & 0.028 & 0.062& 0.000 & 1\\
  & & & & Complete &0.617 & 0.098 & 0.10 & 0.010& 0.949 & 0.99\\
  \\
  & & & 0.30 & RC & -33.961 & 0.064 & 0.064 & 0.023& 0.417 & 0.99\\
  & & & & RSRC & -7.769 & 0.088 & 0.092 & 0.009& 0.910 & 0.987\\
  & & & & GRRC & -1.233 & 0.086 & 0.087 & 0.008& 0.944 & 0.996\\
  & & & & GRN & -1.061 & 0.086 & 0.086 & 0.008& 0.948 & 0.998\\
  & & & & Naive & -66.023 & 0.025 & 0.028 & 0.072& 0.000 & 0.999\\
  & & & & Complete &0.543 & 0.098 & 0.100 & 0.010& 0.950 & 0.99\\
  \\
  & & 1 & 0.15 & RC & -33.862 & 0.074 & 0.073 & 0.024& 0.506 & 0.97\\
  & & & & RSRC & -11.666 & 0.099 & 0.102 & 0.013& 0.899 & 0.958\\
  & & & & GRRC & -1.430 & 0.090 & 0.091 & 0.008& 0.942 & 0.996\\
  & & & & GRN & -1.307 & 0.090 & 0.090 & 0.008& 0.944 & 0.996\\
  & & & & Naive & -69.870 & 0.022 & 0.025 & 0.081& 0.000 & 0.998\\
  & & & & Complete &0.617 & 0.098 & 0.100 & 0.010& 0.954 & 0.99\\
  \\
  & & & 0.30 & RC & -35.737 & 0.075 & 0.074 & 0.026& 0.462 & 0.964\\
  & & & & RSRC & -10.432 & 0.101 & 0.104 & 0.013& 0.905 & 0.959\\
  & & & & GRRC & -1.554 & 0.090 & 0.092 & 0.009& 0.948 & 0.994\\
  & & & & GRN & -1.455 & 0.090 & 0.091 & 0.008& 0.948 & 0.996\\
  & & & & Naive & -73.447 & 0.022 & 0.025 & 0.089& 0.000 & 0.996\\
  & & & & Complete &0.567 & 0.098 & 0.100 & 0.010& 0.952 & 0.99\\
  \end{tabular}
}
\end{table}

\begin{table}[H]
\renewcommand{\thetable}{\arabic{table}}
\caption{Simulation results for $\beta_X=\log 1.5$ under correlated, additive measurement error in the outcome and covariate $X$ with gamma distributed error and 75$\%$ censoring for the true event time. For $2000$ simulated data sets, the bias, average bootstrap standard error (ASE) for the 4 proposed estimators, average model standard error (ASE) for naive and complete case, empirical standard error (ESE), mean squared error (MSE), and 95$\%$ coverage probabilities (CP) are presented.}
\label{correlated_gamma_err_table2}
\centering
\scalebox{0.71}{
  \begin{tabular}{ccccccccccH}
  
  $\beta_X$ & $\sigma^{2}_{\nu}$ & $\sigma^2_{\epsilon}$ & $\sigma_{\nu,\epsilon}$ & Method & $\%$ Bias & ASE & ESE & MSE & CP & Power \\
  \hline
  log(1.5) & & & & True & 0.395 & 0.054 & 0.056 & 0.003 & 0.936 & 1\\
  \\
  & 0.5 & 0.5 & 0.15 & RC & -25.946 & 0.097 & 0.095 & 0.020& 0.807 & 0.878\\
  & & & & RSRC & -7.966 & 0.124 & 0.129 & 0.018& 0.919 & 0.855\\
  & & & & GRRC & 1.110 & 0.148 & 0.146 & 0.021& 0.953 & 0.809\\
  & & & & GRN &1.159 & 0.148 & 0.144 & 0.021& 0.952 & 0.816\\
  & & & & Naive &-68.835 & 0.046 & 0.049 & 0.080& 0.000 & 0.77\\
  & & & & Complete &2.318 & 0.177 & 0.180 & 0.032& 0.950 & 0.658\\
  \\
  & & & 0.30 & RC & -30.582 & 0.097 & 0.096 & 0.025& 0.734 & 0.834\\
  & & & & RSRC & -5.105 & 0.125 & 0.132 & 0.018& 0.920 & 0.879\\
  & & & & GRRC & 1.061 & 0.149 & 0.146 & 0.021& 0.950 & 0.808\\
  & & & & GRN & 1.554 & 0.148 & 0.144 & 0.021& 0.952 & 0.824\\
  & & & & Naive & -79.292 & 0.046 & 0.050 & 0.106& 0.000 & 0.44\\
  & & & & Complete &2.417 & 0.177 & 0.182 & 0.033& 0.947 & 0.652\\
  \\
  & & 1 & 0.15 & RC & -27.154 & 0.111 & 0.110 & 0.024& 0.806 & 0.776\\
  & & & & RSRC & -9.939 & 0.140 & 0.150 & 0.024& 0.912 & 0.764\\
  & & & & GRRC & 0.937 & 0.158 & 0.153 & 0.023& 0.958 & 0.756\\
  & & & & GRN & 1.061 & 0.157 & 0.152 & 0.023& 0.954 & 0.762\\
  & & & & Naive & -75.666 & 0.040 & 0.043 & 0.096& 0.000 & 0.677\\
  & & & & Complete &2.220 & 0.177 & 0.181 & 0.033& 0.947 & 0.654\\
  \\
  & & & 0.30 & RC & -31.470 & 0.110 & 0.109 & 0.028& 0.748 & 0.727\\
  & & & & RSRC & -7.670 & 0.143 & 0.155 & 0.025& 0.908 & 0.774\\
  & & & & GRRC & 0.913 & 0.158 & 0.153 & 0.024& 0.954 & 0.756\\
  & & & & GRN & 1.529 & 0.157 & 0.153 & 0.023& 0.953 & 0.766\\
  & & & & Naive & -83.287 & 0.040 & 0.043 & 0.116& 0.000 & 0.396\\
  & & & & Complete &2.664 & 0.177 & 0.179 & 0.032& 0.952 & 0.652\\
  \\
  & 1 & 0.5 & 0.15 & RC & -25.107 & 0.107 & 0.108 & 0.022& 0.842 & 0.802\\
  & & & & RSRC & -12.110 & 0.138 & 0.149 & 0.025& 0.906 & 0.735\\
  & & & & GRRC & 1.554 & 0.150 & 0.145 & 0.021& 0.954 & 0.82\\
  & & & & GRN &1.603 & 0.149 & 0.144 & 0.021& 0.954 & 0.822\\
  & & & & Naive &-63.088 & 0.046 & 0.050 & 0.068& 0.001 & 0.878\\
  & & & & Complete &2.738 & 0.178 & 0.181 & 0.033& 0.948 & 0.65\\
  \\
  & & & 0.30 & RC & -27.820 & 0.106 & 0.105 & 0.024& 0.810 & 0.786\\
  & & & & RSRC & -8.484 & 0.138 & 0.150 & 0.024& 0.917 & 0.77\\
  & & & & GRRC & 1.159 & 0.150 & 0.149 & 0.022& 0.952 & 0.808\\
  & & & & GRN & 1.332 & 0.149 & 0.147 & 0.022& 0.949 & 0.815\\
  & & & & Naive & -70.413 & 0.046 & 0.049 & 0.084& 0.000 & 0.729\\
  & & & & Complete &2.713 & 0.177 & 0.182 & 0.033& 0.949 & 0.656\\
  \\
  & & 1 & 0.15 & RC & -26.439 & 0.122 & 0.122 & 0.026& 0.836 & 0.695\\
  & & & & RSRC & -14.675 & 0.155 & 0.171 & 0.033& 0.895 & 0.631\\
  & & & & GRRC & 0.715 & 0.158 & 0.152 & 0.023& 0.954 & 0.756\\
  & & & & GRN & 1.061 & 0.157 & 0.152 & 0.023& 0.952 & 0.765\\
  & & & & Naive & -71.128 & 0.040 & 0.042 & 0.085& 0.000 & 0.812\\
  & & & & Complete &2.220 & 0.177 & 0.178 & 0.032& 0.954 & 0.657\\
  \\
  & & & 0.30 & RC & -29.448 & 0.121 & 0.121 & 0.029& 0.810 & 0.666\\
  & & & & RSRC & -11.444 & 0.156 & 0.174 & 0.032& 0.899 & 0.659\\
  & & & & GRRC & 1.208 & 0.160 & 0.154 & 0.024& 0.956 & 0.751\\
  & & & & GRN & 1.455 & 0.158 & 0.152 & 0.023& 0.954 & 0.759\\
  & & & & Naive & -76.801 & 0.040 & 0.043 & 0.099& 0.000 & 0.646\\
  & & & & Complete &3.132 & 0.178 & 0.178 & 0.032& 0.955 & 0.658\\
  \end{tabular}
}
\end{table}

\begin{table}[H]
\renewcommand{\thetable}{\arabic{table}}
\caption{Simulation results for $\beta_X=\log 3$ under correlated, additive measurement error in the outcome and covariate $X$ with gamma distributed error and 25$\%$ censoring for the true event time. For $2000$ simulated data sets, the bias, average bootstrap standard error (ASE) for the 4 proposed estimators, average model standard error (ASE) for naive and complete case, empirical standard error (ESE), mean squared error (MSE), and 95$\%$ coverage probabilities (CP) are presented.}
\label{correlated_gamma_err_table3}
\centering
\scalebox{0.71}{
  \begin{tabular}{ccccccccccH}
  
  $\beta_X$ & $\sigma^{2}_{\nu}$ & $\sigma^2_{\epsilon}$ & $\sigma_{\nu,\epsilon}$ & Method & $\%$ Bias & ASE & ESE & MSE & CP & Power \\
  \hline
  log(3) & & & & True & 0.146 & 0.037 & 0.038 & 0.001 & 0.944 & 1\\
  \\
  & 0.5 & 0.5 & 0.15 & RC & -39.113 & 0.071 & 0.072 & 0.190& 0.002 & 1\\
  & & & & RSRC & -28.864 & 0.089 & 0.094 & 0.109& 0.106 & 1\\
  & & & & GRRC & -0.965 & 0.118 & 0.119 & 0.014& 0.937 & 1\\
  & & & & GRN &-0.901 & 0.118 & 0.119 & 0.014& 0.937 & 1\\
  & & & & Naive &-60.376 & 0.025 & 0.036 & 0.441& 0.000 & 1\\
  & & & & Complete &0.819 & 0.119 & 0.121 & 0.015& 0.948 & 1\\
  \\
  & & & 0.30 & RC & -40.879 & 0.072 & 0.074 & 0.207& 0.002 & 1\\
  & & & & RSRC & -29.710 & 0.093 & 0.099 & 0.116& 0.122 & 1\\
  & & & & GRRC & -1.047 & 0.122 & 0.122 & 0.015& 0.936 & 1\\
  & & & & GRN & -0.947 & 0.120 & 0.121 & 0.015& 0.934 & 1\\
  & & & & Naive & -62.998 & 0.025 & 0.038 & 0.480& 0.000 & 1\\
  & & & & Complete &0.892 & 0.119 & 0.122 & 0.015& 0.948 & 1\\
  \\
  & & 1 & 0.15 & RC & -44.438 & 0.090 & 0.093 & 0.247& 0.018 & 1\\
  & & & & RSRC & -34.726 & 0.106 & 0.114 & 0.159& 0.128 & 1\\
  & & & & GRRC & -1.265 & 0.129 & 0.132 & 0.018& 0.932 & 0.999\\
  & & & & GRN & -1.192 & 0.127 & 0.128 & 0.016& 0.931 & 0.998\\
  & & & & Naive & -71.254 & 0.020 & 0.035 & 0.614& 0.000 & 1\\
  & & & & Complete &0.856 & 0.119 & 0.121 & 0.015& 0.948 & 1\\
  \\
  & & & 0.30 & RC & -45.912 & 0.090 & 0.094 & 0.263& 0.015 & 1\\
  & & & & RSRC & -35.208 & 0.110 & 0.119 & 0.164& 0.145 & 1\\
  & & & & GRRC & -1.338 & 0.131 & 0.131 & 0.017& 0.930 & 0.998\\
  & & & & GRN & -1.320 & 0.128 & 0.129 & 0.017& 0.930 & 1\\
  & & & & Naive & -73.056 & 0.020 & 0.036 & 0.646& 0.000 & 1\\
  & & & & Complete &0.819 & 0.119 & 0.121 & 0.015& 0.947 & 1\\
  \\
  & 1 & 0.5 & 0.15 & RC & -45.066 & 0.074 & 0.074 & 0.251& 0.000 & 1\\
  & & & & RSRC & -32.204 & 0.095 & 0.100 & 0.135& 0.079 & 1\\
  & & & & GRRC & -0.664 & 0.119 & 0.120 & 0.014& 0.941 & 1\\
  & & & & GRN &-0.674 & 0.118 & 0.119 & 0.014& 0.937 & 1\\
  & & & & Naive &-63.871 & 0.025 & 0.034 & 0.494& 0.000 & 1\\
  & & & & Complete &0.874 & 0.119 & 0.122 & 0.015& 0.946 & 1\\
  \\
  & & & 0.30 & RC & -46.322 & 0.074 & 0.074 & 0.264& 0.000 & 1\\
  & & & & RSRC & -32.668 & 0.097 & 0.103 & 0.139& 0.095 & 1\\
  & & & & GRRC & -0.819 & 0.121 & 0.120 & 0.014& 0.938 & 1\\
  & & & & GRN & -0.755 & 0.119 & 0.120 & 0.014& 0.937 & 1\\
  & & & & Naive & -65.883 & 0.025 & 0.035 & 0.525& 0.000 & 1\\
  & & & & Complete &0.847 & 0.119 & 0.121 & 0.015& 0.950 & 1\\
  \\
  & & 1 & 0.15 & RC & -49.171 & 0.091 & 0.093 & 0.300& 0.008 & 1\\
  & & & & RSRC & -36.755 & 0.112 & 0.118 & 0.177& 0.124 & 1\\
  & & & & GRRC & -0.992 & 0.126 & 0.127 & 0.016& 0.938 & 1\\
  & & & & GRN & -0.956 & 0.125 & 0.126 & 0.016& 0.937 & 1\\
  & & & & Naive & -73.393 & 0.020 & 0.033 & 0.651& 0.000 & 1\\
  & & & & Complete &0.828 & 0.119 & 0.122 & 0.015& 0.949 & 1\\
  \\
  & & & 0.30 & RC & -50.254 & 0.091 & 0.093 & 0.314& 0.006 & 1\\
  & & & & RSRC & -37.029 & 0.114 & 0.122 & 0.180& 0.130 & 1\\
  & & & & GRRC & -1.001 & 0.128 & 0.128 & 0.016& 0.936 & 1\\
  & & & & GRN & -0.956 & 0.126 & 0.128 & 0.016& 0.935 & 0.998\\
  & & & & Naive & -74.831 & 0.020 & 0.034 & 0.677& 0.000 & 1\\
  & & & & Complete &0.856 & 0.119 & 0.121 & 0.015& 0.949 & 1\\
  \\
  \end{tabular}
}
\end{table}

\section*{Appendix F: Misclassification table}

\begin{table}[H]
\renewcommand{\thetable}{\arabic{table}}
\caption{Simulation results for $\beta_X=\log 1.5$ under misspecification and correlated, additive measurement error in the outcome and covariate $X$ with normally distributed error, 75$\%$ censoring for the true event time, $90\%$ sensitivity, and $90\%$ specificity. For $2000$ simulated data sets, the bias, average bootstrap standard error (ASE) for the 4 proposed estimators, average model standard error (ASE) for naive and complete case, empirical standard error (ESE), mean squared error (MSE), and 95$\%$ coverage probabilities (CP) are presented.}
\label{misclassification_table}
\centering
\scalebox{0.90}{
  \begin{tabular}{ccccccccccH}
  
  $\beta_X$ & $\sigma^{2}_{\nu}$ & $\sigma^2_{\epsilon}$ & $\sigma_{\nu,\epsilon}$ & Method & $\%$ Bias & ASE & ESE & MSE & CP & Power \\
  \hline
  log(1.5) & & & & True & -0.099 & 0.055 & 0.054 & 0.003 & 0.953 & 1\\
  \\
  & 0.5 & 0.5 & 0.15 & RC & -43.111 & 0.106 & 0.101 & 0.041 & 0.611 & 0.599\\
  & & & & RSRC & -40.842 & 0.118 & 0.117 & 0.041 & 0.681 & 0.539\\
  & & & & GRRC & -0.049 & 0.170 & 0.163 & 0.027 & 0.952 & 0.679\\
  & & & & GRN & 0.641 & 0.172 & 0.164 & 0.027 & 0.954 & 0.684\\
  & & & & Naive &-141.097 & 0.042 & 0.042 & 0.329 & 0.000 & 0.978\\
  & & & & Complete & -0.025 & 0.177 & 0.178 & 0.032 & 0.953 & 0.635\\
\end{tabular}
}
\end{table}

\section*{Appendix G: VCCC eligibility criteria}

We analyzed data on 4797 HIV-positive patients that established care at the VCCC between 1998 and 2013. For the virologic failure outcome, patients were excluded if they had an indeterminate ART start date, started ART prior to enrollment, had no CD4 count measurement between 180 days before or 30 days after starting ART, or had no follow-up after starting ART. Using the unvalidated data, 2143 patients met the criteria for inclusion, of which 1863 met the criteria using the validated data. These 1863 patients were used in all further analyses to ensure that any differences between estimators are not due to the differences in included patients. For the ADE outcome, the exclusion criteria was similar to that of the former analysis except we additionally excluded patients that had an ADE before ART initiation and those with indeterminate ADE dates. Using the unvalidated data, 1995 patients met the ADE analysis criteria, of which 1595 met the criteria using the validated data. Again, these 1595 were used in all further ADE analyses. Note that for both analyses, failures within 6 months of ART start were not considered a true failure due to the time required by the regimen to be efficacious. In addition, we made some further simplifying assumptions for the purpose of this data example for ease of exposition. Specifically, we removed subjects from the analyses that were not in both the unvalidated and validated datasets for ease of interpretation and selected validation subsets as if we did not validate all subjects. This was done to highlight the application of our methods and be able to effectively compare their relative performance.

Of the 1863 patients in the analysis of the virologic failure outcome, 20 were incorrectly classified as having failed, resulting in a $1\%$ misclassification rate. There were 386 incorrectly recorded event times, with the error having mean and standard deviation of $-0.13$ and $1.1$ years, respectively. CD4 count at ART start was incorrect for 125 patients, with the error having mean and standard deviation of $21$ and $164$ $\text{cell}/\text{mm}^3$, respectively. The correlation between the error in the failure times and CD4 count at ART initiation for subjects with both types of error was $-0.17$.

Of the 1595 patients in the analysis of the ADE outcome, 161 were incorrectly classified as having had an ADE and 12 were incorrectly classified as having been censored, resulting in an appreciable misclassification rate of $11\%$. There were 551 incorrectly recorded event times, with the error having mean and standard deviation of $-0.75$ and $2.89$ years, respectively. CD4 count at ART start was incorrect for 107 patients, with the error having mean and standard deviation of $10$ and $154$ $\text{cell}/\text{mm}^3$, respectively. The correlation between the error in the failure times and CD4 count at ART initiation for subjects with both types of error was $-0.10$.

\clearpage 

\section*{Appendix H: VCCC table}

\begin{table}[H]
\renewcommand{\thetable}{\arabic{table}}
\caption{The hazard ratios (HR) and their corresponding $95\%$ confidence intervals (CI) for a 100 cell/$\text{mm}^3$ increase in CD4 count at ART initiation and 10 year increase in age at CD4 count measurement. The CIs are calculated using the bootstrap for the RC, RSRC, GRRC, and GRN estimators.}
\label{vccc_table}
\centering
\scalebox{0.90}{
\begin{tabular}{cccc}
Outcome & Method & $100 \times $ CD4 & $10 \times $ Age \\
\hline 
Time to virologic failure & True & 0.902 (0.869, 0.935) & 0.860 (0.806, 0.916) \\
& RC & 0.920 (0.888, 0.953) & 0.880 (0.825, 0.939) \\
& RSRC & 0.918 (0.885, 0.953) & 0.879 (0.821, 0.942) \\
& GRRC & 0.918 (0.883, 0.954) & 0.869 (0.811, 0.932) \\
& GRN & 0.918 (0.882, 0.956) & 0.869 (0.802, 0.942) \\
& Naive & 0.918 (0.885, 0.953) & 0.878 (0.824, 0.936) \\
& HT & 0.929 (0.852, 1.012) & 0.790 (0.679, 0.919)\\
\\
Time to ADE & True & 0.693 (0.593, 0.809) & 0.829 (0.671, 1.023) \\
& RC & 0.899 (0.832, 0.971) & 1.071 (0.940, 1.221) \\
& RSRC & 0.895 (0.827, 0.969) & 1.073 (0.938, 1.226) \\
& GRRC & 0.694 (0.565, 0.852) & 0.883 (0.632, 1.234) \\
& GRN & 0.693 (0.564, 0.853) & 0.883 (0.622, 1.253) \\
& Naive & 0.910 (0.841, 0.986) & 1.087 (0.957, 1.235) \\
& HT & 0.748 (0.597, 0.939) & 1.114 (0.757, 1.640) \\
\end{tabular}
}
\end{table}

\section*{Appendix I: Example \textsf{R} code}

The code below demonstrates how to implement the Regression Calibration and Generalized Raking Naive methods for example datasets. This example assumes the validation subset was selected as a simple random sample and that there are two covariates (X is error prone and Z is error free). Note that the code only demonstrates how to obtain estimates; standard errors must be calculated using the stratified bootstrap as described in the paper. Full code implementing all methods discussed in the paper (including standard errors) is available at https://github.com/ericoh17/RRCME. 

\begin{verbatim}
library(dplyr)
library(survival)
library(survey)

# Example datasets
full_dat <- read.csv("example_dat.csv", row.names = 1)
valid_subset <- read.csv("example_valid_subset.csv", row.names = 1)

full_dat$time <- full_dat$delta <- full_dat$x <- NA

full_dat$time[full_dat$randomized == TRUE] <- valid_subset$time
full_dat$delta[full_dat$randomized == TRUE] <- valid_subset$delta
full_dat$x[full_dat$randomized == TRUE] <- valid_subset$x

### Regression Calibration ###
# Calibrate the covariate
x_calib_model <- lm(x ~ x_star + z, data = valid_subset)
x_hat <- predict(x_calib_model, data = full_dat)

# Calibrate the outcome
w_calib_model <- lm(total_y_err ~ x_star + z, data = valid_subset)
w_hat <- predict(w_calib_model, data = full_dat)
time_hat <- full_dat$time_star - w_hat

# Fit RC model
rc_mod <- coxph(Surv(time_hat, full_dat$delta_star) ~ x_hat + full_dat$z)

# Extract RC coefficients
beta_x_RC <- rc_mod$coef[1]
beta_z_RC <- rc_mod$coef[2]

### Generalized Raking Naive ###
# Fit naive model
naive_mod <- coxph(Surv(time_star, delta_star) ~ x_star + z, data = full_dat)

# Extract influence functions from naive model
IF_naive <- data.frame(resid(naive_mod, "dfbeta"))
colnames(IF_naive) <- paste("if", 1:2, sep = "")
full_IF_dat <- dplyr::bind_cols(full_dat, IF_naive)

# Calculate raking weights
IF_design <- twophase(id = list(~id, ~id), subset = ~randomized, 
                      data = full_IF_dat)
IF_raking <- calibrate(IF_design, phase = 2, formula = ~if1+if2, 
                       calfun = "raking")

# Fit raking model
raking_mod <- svycoxph(Surv(time, delta) ~ x + z, design = IF_raking)

# Extract raking coefficients
beta_x_GRN <- raking_mod$coef[1]
beta_z_GRN <- raking_mod$coef[2]
\end{verbatim}

\end{document}